\ifdefined\pdfminorversion\pdfminorversion=7\fi
\documentclass[11pt,letterpaper]{article}

\usepackage[T1]{fontenc}
\usepackage[utf8]{inputenc}
\usepackage[english]{babel}

\usepackage{tgpagella}

\usepackage{microtype}

\usepackage{amsmath,amssymb,amsthm,mathtools}
\usepackage{bm}
\usepackage{setspace}
\usepackage{needspace}

\usepackage[
  letterpaper,
  margin=68bp,
  footskip=29bp
]{geometry}

\usepackage{graphicx}
\usepackage{booktabs}
\usepackage{threeparttable}
\usepackage{array}
\usepackage{siunitx}
\usepackage{caption}
\usepackage{float}
\usepackage{flafter}
\usepackage[section]{placeins}
\newcommand{\figurenote}[1]{\par\medskip
  \begin{minipage}{\linewidth}
    \small\singlespacing\noindent#1\par
  \end{minipage}\par
}

\usepackage{enumitem}
\setlist{nosep,leftmargin=1.5em}

\usepackage{titlesec}
\titleformat{\section}[block]
  {\Large\bfseries}
  {\thesection}{1em}{}
\titleformat{\subsection}
  {\large\bfseries}
  {\thesubsection}{1em}{}
\titleformat{\subsubsection}[block]
  {\normalsize\bfseries}
  {\thesubsubsection}{1em}{}

\titlespacing*{\section}{0pt}{3.5ex plus 1ex minus .2ex}{2.3ex plus .2ex}
\titlespacing*{\subsection}{0pt}{3.25ex plus 1ex minus .2ex}{1.5ex plus .2ex}
\titlespacing*{\subsubsection}{0pt}{3.25ex plus 1ex minus .2ex}{1.5ex plus .2ex}

\usepackage{titling}
\pretitle{\begin{center}\LARGE}
\posttitle{\par\end{center}\vspace{-0.5em}}
\preauthor{\begin{center}\large}
\postauthor{\par\end{center}\vspace{-0.7em}}
\predate{\begin{center}\large}
\postdate{\par\end{center}}

\renewenvironment{abstract}{
  \small\singlespacing
  {\centering\bfseries Abstract\par}
  \smallskip
  \noindent\ignorespaces
}{
  \par
}

\usepackage[round,authoryear]{natbib}
\usepackage[bottom,hang,flushmargin]{footmisc}
\usepackage{xcolor}
\definecolor{citeblue}{cmyk}{1,1,0,0}
\definecolor{brickred}{cmyk}{0,0.89,0.94,0.28}

\usepackage[
  colorlinks=true,
  hyperfootnotes=false,
  linkcolor=brickred,
  citecolor=citeblue,
  urlcolor=brickred,
  pdfborder={0 0 0}
]{hyperref}
\hypersetup{
  pdftitle={Probabilistic Causal Panel Analysis},
  pdfauthor={Carl Bonander, Qinyun Lin, Ulf Str\"omberg, and Marta Blangiardo},
  pdfkeywords={causal inference, panel data, Bayesian hierarchical models, counterfactual imputation, synthetic control}
}
\usepackage[capitalize,noabbrev]{cleveref}

\newtheoremstyle{paperstyle}
  {8pt}
  {8pt}
  {\itshape}
  {}
  {\bfseries}
  {.}
  {0.5em}
  {}
\theoremstyle{paperstyle}
\newtheorem{theorem}{Theorem}[section]
\newtheorem{proposition}{Proposition}[section]

\theoremstyle{definition}
\newtheorem{assumption}{Assumption}[section]

\theoremstyle{paperstyle}

\newcommand{\E}{\mathbb{E}}
\newcommand{\Var}{\operatorname{Var}}
\newcommand{\N}{\mathcal{N}}
\newcommand{\trans}{^{\mathsf T}}

\newcommand{\vect}[1]{\bm{#1}}
\newcommand{\mat}[1]{\bm{#1}}

\crefname{assumption}{assumption}{assumptions}
\Crefname{assumption}{Assumption}{Assumptions}

\title{Probabilistic Causal Panel Analysis}
\author{Carl Bonander, Qinyun Lin, Ulf Str\"omberg and Marta Blangiardo\thanks{Bonander: Karlstad Business School and the Center for Societal Risk Research, Karlstad University, Karlstad, Sweden. Lin and Str\"omberg: School of Public Health and Community Medicine, University of Gothenburg, Sweden. Blangiardo: MRC Centre for Environment and Health, Department of Epidemiology and Biostatistics, School of Public Health, Imperial College London, UK. Correspondence: \href{mailto:carl.bonander@kau.se}{carl.bonander@kau.se}.}}
\date{November 14, 2026}
\begin{document}
\addtocontents{toc}{\protect\setcounter{tocdepth}{-1}}

\pagestyle{plain}
\maketitle

\vspace{1.4em}
\begin{center}
\begin{minipage}{0.885\textwidth}
\begin{abstract}

Many causal panel estimators construct counterfactuals using weights or latent structure estimated directly from observed outcomes, making their estimates sensitive to observation noise. Bayesian hierarchical models can account for this noise and pool information across related observations. Existing Bayesian approaches to causal panel analysis, however, typically specify counterfactual imputation within the probabilistic model, so changing estimators may require a new formulation. We develop a framework that learns the latent untreated panel with a probabilistic model and applies established causal estimators to its posterior draws. The model accommodates non-Gaussian observations and structured pooling, while re-estimation across draws propagates uncertainty through counterfactual imputation. We establish conditions for posterior consistency and characterize how pooling affects imputation risk. Under correct posterior specification, we show that the excess Bayes risk of our approach is bounded by the causal estimator's expected squared imputation error when given the true latent outcomes in observed untreated cells. We also develop a fast cross-validation approach for model selection and calibration of predictive uncertainty for treatment effects. We use simulations to examine when pooling helps or harms imputation and cross-validation to assess whether probabilistic modeling improves accuracy in multiple empirical applications.

\end{abstract}

  \vspace{1.7em}
\end{minipage}
\end{center}

\clearpage
\setstretch{1.2134}

\section{Introduction}
\label{sec:pcpa-introduction}

Panel data are widely used to estimate the effects of policies introduced in some units and periods but not others. When adoption is not random, these evaluations require a credible comparison with what would have happened in the treated units without the policy \citep{athey2017state,arkhangelsky2024causal}. Difference-in-differences makes the parallel-trends assumption that, without treatment, average outcomes would have changed by the same amount in treated and comparison units \citep{borusyak2024revisiting,liu2024practical}. This assumption can fail when unobserved influences change over time and affect units differently. Synthetic control and factor-based methods address some of these departures by using pretreatment histories and outcomes across units to estimate counterfactual trajectories \citep{abadie2010synthetic,bai2009panel,xu2017generalized,athey2021matrix,choi2024matrix,arkhangelsky2021synthetic}.

The weights or latent factors used to construct these counterfactuals are generally estimated directly from the observed outcome panel. Yet the observations may be event counts, estimated rates or proportions, or survey-based means, with different amounts of information across units and periods. Observation noise can obscure persistent patterns in outcome trajectories, distorting the estimated weights or factors \citep{abadie2021using,amjad2018robust,fan2016projected}. The resulting errors in counterfactual imputation then carry through to the estimated treatment effects. An outcome model that accounts for measurement precision and dependence across observations offers a way to improve the inputs to causal estimation.

Bayesian hierarchical models provide these capabilities through observation likelihoods and pooling across related units and periods \citep{rao2015small,knorrheld2000bayesian,rue2009approximate,riebler2016intuitive}. Existing Bayesian approaches to causal panel analysis also specify counterfactual extrapolation within the probabilistic model \citep[e.g.,][]{brodersen2015inferring,tanaka2021bayesian,pang2022bayesian}. Adopting a different causal estimator can therefore require a new joint formulation and estimation algorithm. Applied researchers would benefit from being able to combine flexible outcome models and structured pooling with established causal estimators without rebuilding the Bayesian analysis around each one.

We develop a framework that combines Bayesian learning of the latent untreated panel with counterfactual imputation by established causal estimators. A hierarchical model learns the latent process from observed untreated outcomes, accounting for observation noise and pooling information across related cells. The causal estimator is then applied to joint posterior draws, allowing uncertainty about the latent untreated process to propagate through its fitted weights or factors to the counterfactuals. Its estimation procedure remains unchanged, with identifying assumptions applied on the latent scale. For inference on realized treatment effects, we also use information from the probabilistic model about variation around these imputations to construct calibrated predictive distributions.

Causal estimation in this framework requires no estimator-specific Bayesian sampler. The construction can use joint posterior draws obtained by Markov chain Monte Carlo, including for non-Gaussian latent models. For latent Gaussian models, the framework can be implemented as post-processing of models fitted with integrated nested Laplace approximations (INLA; \citealp{rue2009approximate}). Researchers can therefore combine established causal estimators with scalable Bayesian models widely used in disease mapping and spatiotemporal modeling. The same probabilistic fit supports different causal estimators and tuning choices. Building on leave-group-out methods \citep{liu2025leavegroup,adin2024automatic}, we also develop fast cross-validation for model selection and predictive calibration in latent Gaussian models, without refitting the probabilistic model for each validation fold.

The benefit of probabilistic modeling depends on how accurately we learn the latent untreated panel and how the causal estimator responds to errors in that learning. We establish conditions under which posterior concentration carries over to counterfactual imputation and characterize how the estimator's sensitivity determines the balance between smoothing bias and variance reduction. Under a correctly specified posterior for the latent outcomes in observed untreated cells, we also show that the excess Bayes risk of our approach is bounded by the expected squared imputation error of the causal estimator when given the true latent outcomes. If the causal estimator recovers the missing latent conditional mean from those outcomes, our approach attains the Bayes optimum, which means that no estimator applied directly to observed outcomes can achieve lower expected squared imputation error.

We next discuss the related literature. Section~\ref{sec:pcpa-setup} then introduces the formal setup and probabilistic models. In Section~\ref{sec:pcpa-method}, we develop the proposed method, including predictive inference and cross-validation. Section~\ref{sec:pcpa-theory} presents the theoretical results, while Section~\ref{sec:pcpa-evidence} examines the framework through numerical experiments and empirical case studies. Section~\ref{sec:pcpa-discussion} concludes.

\subsection{Related literature}
\label{sec:pcpa-related}

Most causal panel estimators can be written as imputation estimators: they predict the missing untreated outcomes of treated units from observed untreated outcomes and estimate effects by comparing those predictions with the treated units' observed outcomes \citep{borusyak2024revisiting,liu2024practical}. The estimator's identifying assumptions determine how this prediction extends from observed to missing cells. In their standard forms, difference-in-differences, synthetic controls, and matrix completion impute directly from observed outcomes, under assumptions such as parallel trends or the requirement that a weighted combination of control units tracks the treated unit after treatment as it did before. Bayesian approaches impute from a joint probabilistic model for outcomes and counterfactuals \citep{brodersen2015inferring,kim2020bayesian,pang2022bayesian,tanaka2021bayesian,ben2023estimating,samartsidis2024bayesian}. These models can describe how outcomes are measured, including non-Gaussian observations and structured dependence \citep{bonander2026spatial,vega2025spatio,grossi2026spatial}, but the probabilistic specification also determines how counterfactuals are extrapolated. Adopting a different causal estimator can therefore require another joint formulation. Work on Bayesian missing-data and sensitivity analysis has proposed specifying the observed-data model separately from the assumptions used to extrapolate to missing values: the former can be checked against observed data, whereas the latter are not identified by those data \citep{daniels2008missing,linero2018bayesian,franks2020flexible}. We adopt the same separation. The probabilistic model learns the latent untreated process from observed outcomes, while the causal estimator supplies the extrapolation. Different causal estimators and identifying assumptions can therefore be examined with the probabilistic fit held fixed. These identifying assumptions apply to the latent untreated process rather than directly to the noisy observed outcomes.

Noise in the observed outcomes is a separate problem for established estimators, which fit their weights or latent structure directly to the observed panel. In synthetic controls, the weights are chosen to reproduce the treated unit's pretreatment outcomes. When those outcomes contain transitory noise, the fitted weights need not reproduce the persistent structure needed for counterfactual prediction. With a fixed donor pool, \citet{ferman2021synthetic} show that treatment-effect estimates can remain biased even as the number of pretreatment periods grows. Proximal synthetic controls address this by changing the estimator. They treat control outcomes as noisy proxies for the latent factors behind untreated outcomes and identify the weights through conditions that link the proxies to those factors \citep{qiu2024doubly,shi2026proximal}. This requires assumptions on the proxies rather than a probabilistic model for observation noise. \citet{amjad2018robust} instead denoise the outcome matrix with a low-rank approximation before constructing a synthetic control. Our framework extends this approach to causal panel estimators beyond synthetic controls and uses an observation model that accounts for differences in precision across cells. Posterior draws also carry uncertainty about the latent inputs through counterfactual imputation, which a single denoised matrix cannot. Conventional factor and matrix-completion estimators already reduce noise by exploiting low-rank structure. Low rank alone, however, does not encode geographic adjacency or temporal proximity. Additional information about the latent process can improve recovery of the factor structure \citep{fan2016projected}. Our framework allows researchers to use spatial neighborhoods and temporal dependence for this purpose through the probabilistic model, without modifying the causal estimator.

Applying an estimator to each posterior draw connects our construction to inference based on transformed posteriors \citep{wang2023coverage}, including posterior projection \citep{astfalck2026posterior}, where draws from an unconstrained model are projected onto a constraint set. Euclidean projection onto a closed convex set is nonexpansive, meaning that it never moves two draws further apart. Posterior concentration around a truth in that set then carries over to the projected draws. Causal estimators need not have a comparable property, and small errors in their inputs can be amplified in the imputed counterfactuals. The theory in Section~\ref{sec:pcpa-theory} therefore ties the transfer of concentration to recovery and stability of the causal estimator, and characterizes how its sensitivity to input error changes the balance between smoothing bias and variance.

With one or a few treated units, a common target is the realized treatment effect, defined as the difference between a treated unit's observed outcome and its missing untreated outcome rather than as an average over a population of units. Inference then amounts to predicting the missing outcomes. Prediction intervals for synthetic controls \citep{cattaneo2021prediction,cattaneo2025staggered} and conformal inference \citep{chernozhukov2021conformal} give guarantees under conditions on estimation and prediction errors, including their dependence. Joint Bayesian models give predictive distributions directly, valid to the extent that the full specification is credible \citep{pang2022bayesian}. Our predictive construction draws on both. As in \citet{cattaneo2021prediction}, we separate uncertainty in constructing the counterfactual from variation around it, but we describe that variation with the probabilistic model of the latent process and the observation mechanism. For an aggregate treatment effect, that variation depends on cross-cell dependence as well as marginal uncertainty. Calibration of individual predictive intervals therefore need not calibrate the aggregate. Drawing on forecast postprocessing \citep{sansom2016best,gneiting2007strictly}, we develop a common calibration procedure for the causal estimators and probabilistic models in our framework. We score predictive distributions of placebo effects formed from withheld untreated outcomes against zero, using the weighting and aggregation of the treatment effect of interest. Calibration thus evaluates uncertainty for the causal estimand, including the contribution of dependence across missing outcomes.

\section{Setup and probabilistic models}
\label{sec:pcpa-setup}
\label{sec:setup}

\subsection{Treatment and estimand}
\label{subsec:pcpa-target}
\label{subsec:causal-setup}

We observe units \(i=1,\ldots,N\) over periods \(t=1,\ldots,T\). Let \(A_{it}\in\{0,1\}\) indicate whether unit \(i\) is treated in period \(t\), and collect the realized assignments in \(\mat A=(A_{it})\). Let \(Y_{it}\) denote the observed outcome and \(Y_{it}(0)\) the potential outcome under no treatment \citep{rubin1974estimating,imbens2015causal}. For treated cells, write \(Y_{it}(1)\) for the potential outcome under the realized treatment pattern. The units, periods, and treatment pattern are treated as fixed throughout. The treatment pattern may involve common adoption, staggered adoption, or a more irregular configuration.

Write
\[
\mathcal M=\{(i,t):A_{it}=1\},
\qquad
\Omega=\{(i,t):A_{it}=0\}.
\]
We use observations in $\Omega$ to learn the untreated process under the following assumption.

\begin{assumption}[Consistency on untreated cells]
\label{ass:pcpa-untreated-consistency}
The observed outcomes satisfy
\begin{equation}
Y_{it}=Y_{it}(0),\qquad (i,t)\in\Omega.
\label{eq:pcpa-observed-untreated}
\end{equation}
\label{eq:untreated-consistency}
\end{assumption}

For cells in $\Omega$, this assumption excludes effects before treatment adoption (anticipation) and effects transmitted from treated units (spillovers).\footnote{If anticipation or spillover exposure can be identified, $A_{it}$ can be redefined to include the affected cells in $\mathcal M$. Their effects $Y_{it}-Y_{it}(0)$ can then be estimated if the remaining observations support counterfactual identification \citep[see, e.g.,][]{butts2023difference}.}

Let \(\vect q=\{q_{it}:(i,t)\in\mathcal M\}\) contain fixed nonnegative weights satisfying \(\sum_{(i,t)\in\mathcal M}q_{it}=1\). We target the weighted effect on the realized outcomes in the treated cells,
\begin{equation}
\begin{aligned}
\tau(\vect q)
&= \sum_{(i,t)\in\mathcal M}q_{it}
   \left\{Y_{it}-Y_{it}(0)\right\} \\
&= \sum_{(i,t)\in\mathcal M}q_{it}
   \left\{Y_{it}(1)-Y_{it}(0)\right\}.
\end{aligned}
\label{eq:pcpa-effect-target}
\end{equation}
\label{eq:causal-estimand}
The second equality uses \(Y_{it}=Y_{it}(1)\) on \(\mathcal M\). Under uniform weights, \(\tau(\vect q)\) is the average effect over the realized treated cells and corresponds to the sample average treatment effect on the treated (SATT). Event-time or cohort-specific effects follow by placing positive weight only on the relevant treated cells.

For this target, the treated outcomes are observed, so inference reduces to predicting their missing untreated counterparts on the same outcome scale, including observation variation.\footnote{Effects on underlying mean outcomes are different targets and would require inference for the treated means as well as their untreated counterfactuals.}

\subsection{Probabilistic models for untreated outcomes}
\label{subsec:pcpa-latent-model}
\label{subsec:probabilistic-panel}

We distinguish the latent untreated process from the observation process through which it is measured. Let $\mathbf Z(0)$ denote the latent panel and let $\boldsymbol\theta$ collect the remaining likelihood and latent-field parameters. We assume the following conditional structure for the observation process.
\begin{assumption}[Conditional observation model]
\label{ass:pcpa-conditional-observations}
Under the probabilistic model, the untreated potential outcomes are independent across cells conditional on $\mathbf Z(0)$ and $\boldsymbol\theta$. Their conditional laws satisfy

\begin{equation}
Y_{it}(0)\mid\mathbf Z(0),\boldsymbol\theta
\sim p_{it}\{\cdot\mid Z_{it}(0),\boldsymbol\theta\}.
\label{eq:pcpa-observation-model}
\end{equation}
\label{eq:observation-model}
\end{assumption}

Dependence across cells can still arise through $\mathbf Z(0)$ and $\boldsymbol\theta$. The observation model also specifies how the latent process relates to observed outcomes. For example, with Poisson counts and population exposure, $Z_{it}(0)$ can represent the latent log rate. Since the causal estimators in Section~\ref{sec:pcpa-method} are applied to posterior draws of $\mathbf Z_\Omega(0)$, their identifying assumptions must hold on the corresponding latent scale.

To pool information across cells when estimating $\mathbf Z(0)$, we specify
\begin{equation}
\mathbf Z(0)\mid\boldsymbol\theta\sim P_{\boldsymbol\theta},
\qquad \boldsymbol\theta\sim\pi.
\label{eq:pcpa-latent-model}
\end{equation}
\label{eq:latent-panel-model}
The distribution $P_{\boldsymbol\theta}$ can represent spatial, temporal, or other dependence. Under Assumptions~\ref{ass:pcpa-untreated-consistency} and~\ref{ass:pcpa-conditional-observations}, the likelihood for the observed untreated outcomes factors over $\Omega$ according to Equation~\eqref{eq:pcpa-observation-model}. Together with Equation~\eqref{eq:pcpa-latent-model}, this defines the probabilistic model. We use the subscript $P$ for its probabilities and expectations, and write $\mathcal D_\Omega$ for the untreated outcomes and the covariates, exposures, or other quantities entering the model.

Latent Gaussian models offer a broad class of probabilistic specifications in which the observation likelihood and the dependence structure of the latent process can be modeled separately \citep{rue2009approximate}. A useful example is a separable spatiotemporal model that combines autoregressive temporal dependence with a spatial effect mixing graph-structured and independent variation, as in the Besag--York--Molli\'e construction \citep{besag1991bayesian,riebler2016intuitive}. Such models allow information to be pooled across neighboring units and successive periods, while the spatial mixing parameter and temporal correlation admit independence in either dimension as limiting cases.

\section{Probabilistic causal panel analysis}
\label{sec:pcpa-method}
\label{sec:probabilistic-imputation}

We now turn from learning the latent untreated panel to causal extrapolation. We focus on estimators that impute missing untreated outcomes from observed untreated cells, and apply them to the corresponding latent inputs.

\subsection{Counterfactual imputation across posterior draws}
\label{subsec:pcpa-draw-wise}
\label{subsec:posterior-imputation}

Let $k$ index a fully specified causal panel estimator, including its tuning parameters and a deterministic rule for resolving multiple optimizer solutions when needed. For a latent input panel $\mathbf z_\Omega$, write $c_k(\mathbf z_\Omega)\in\mathbb R^{|\mathcal M|}$ for the corresponding counterfactual imputation, and assume that $c_k$ is measurable. Valid imputation relies on the identifying assumptions of the chosen estimator, applied on the latent scale. For example, difference-in-differences applied to latent log rates requires parallel trends in untreated log rates rather than observed rates.

Conditioning the probabilistic model on $\mathcal D_\Omega$ gives a joint posterior for the latent inputs and model parameters. For draw $s$, we obtain
\begin{equation}
\begin{aligned}
\left(\mathbf Z_\Omega^{(s)}(0),\boldsymbol\theta^{(s)}\right)
&\sim\Pi_P(\cdot\mid \mathcal D_\Omega),\\
\widehat{\mathbf Z}_{k,\mathcal M}^{(s)}(0)
&=c_k\!\left\{\mathbf Z_\Omega^{(s)}(0)\right\}.
\end{aligned}
\label{eq:pcpa-draw-wise}
\end{equation}
\label{eq:first-stage-posterior}
\label{eq:drawwise-imputation}
For fixed $k$, applying $c_k$ to the joint posterior draws gives a posterior distribution over counterfactual imputations, formally the pushforward of the posterior for the latent inputs under $c_k$ \citep{butler2018combining}.

For concreteness, consider the synthetic control method \citep{abadie2003economic,abadie2010synthetic}. For a treated unit $i$, never-treated controls $\mathcal C$, and pretreatment periods $\mathcal P_i$, the control weights are chosen to match the treated unit's pretreatment outcomes. For each posterior draw,
\begin{equation}
\begin{aligned}
\widehat{\boldsymbol\omega}_i^{(s)}
&\in\arg\min_{\substack{\omega_j\geq0\\\sum_{j\in\mathcal C}\omega_j=1}}
\sum_{t\in\mathcal P_i}
\left\{Z_{it}^{(s)}(0)-\sum_{j\in\mathcal C}\omega_jZ_{jt}^{(s)}(0)\right\}^2,\\
\widehat Z_{k,it}^{(s)}(0)
&=\sum_{j\in\mathcal C}\widehat\omega_{ij}^{(s)}Z_{jt}^{(s)}(0),
\qquad (i,t)\in\mathcal M.
\end{aligned}
\label{eq:pcpa-synthetic-control}
\end{equation}
Re-estimating the control weights for each draw carries uncertainty in the latent inputs through both the weights and the control outcomes used for imputation. Fixed covariates and other matching predictors can also enter the optimization. Appendix~\ref{app:pcpa-estimators} gives implementations for interactive fixed effects, matrix completion, and ridge-augmented synthetic controls. It also describes a fast algorithm for repeated synthetic-control fits.

\subsection{Posterior predictive inference}
\label{subsec:pcpa-prediction}
\label{subsec:posterior-predictive}

The imputations in Section~\ref{subsec:pcpa-draw-wise} vary across draws because the latent inputs are uncertain. For inference on $\tau(\mathbf q)$, prediction must also account for variation in the missing untreated outcomes around those imputations \citep{cattaneo2021prediction}. Following \citet{pang2022bayesian}, we predict these outcomes conditional on $\mathcal D_\Omega$ and use the observed treated outcomes only to form the effect contrast. On the latent scale, we represent the remaining variation by innovations, as in time-series and state-space models \citep{west1997bayesian,durbin2012time},
\begin{equation}
\begin{aligned}
\mathbf Z_{k,\mathcal M}^{(s)}(0)
&=\widehat{\mathbf Z}_{k,\mathcal M}^{(s)}(0)
  +\mathbf u_{\mathcal M}^{(s)},\\
\mathbb E\!\left\{\mathbf u_{\mathcal M}^{(s)}
\mid\mathbf Z_\Omega^{(s)}(0),\boldsymbol\theta^{(s)}\right\}
&=\mathbf0.
\end{aligned}
\label{eq:pcpa-latent-innovation}
\end{equation}
We center the innovations at zero so that $c_k(\mathbf z_\Omega)$ is the conditional mean under the predictive construction. To describe variation around this imputation, we use the probabilistic model's conditional covariance as a reference. Write
\begin{equation}
\mathbf V_P(\boldsymbol\theta)
=\operatorname{Var}_P\!\left\{\mathbf Z_{\mathcal M}(0)
\mid\mathbf Z_\Omega(0),\boldsymbol\theta\right\},
\qquad
\mathbf D_P(\boldsymbol\theta)
=\operatorname{diag}\!\left\{\mathbf V_P(\boldsymbol\theta)\right\},
\label{eq:pcpa-reference-covariance}
\end{equation}
where $\mathbf D_P(\boldsymbol\theta)$ is the diagonal matrix of conditional variances.\footnote{The notation suppresses possible dependence on the conditioned values of $\mathbf Z_\Omega(0)$. When $\mathbf Z(0)\mid\boldsymbol\theta$ is jointly Gaussian, the conditional covariance depends on the model parameters and the conditioning pattern, but not on the realized latent values. Appendix~\ref{app:pcpa-reference-sampling} provides computational details and the corresponding innovation sampler.} Following work on predictive dispersion in forecasting \citep[see, e.g.,][]{moeller2020probabilistic}, we allow this reference covariance to be calibrated through scale adjustment and correlation shrinkage,
\begin{equation}
\mathbf K_P(\kappa,\chi;\boldsymbol\theta)
=\kappa^2\left[(1-\chi)\mathbf D_P(\boldsymbol\theta)
+\chi\mathbf V_P(\boldsymbol\theta)\right],
\qquad \kappa\geq0,\quad0\leq\chi\leq1.
\label{eq:pcpa-covariance-family}
\end{equation}
The parameter $\kappa$ rescales marginal standard deviations to accommodate overstatement or understatement of the remaining latent variation. For fixed $\kappa$, decreasing $\chi$ from one to zero attenuates the off-diagonal entries while preserving the diagonal $\kappa^2\mathbf D_P(\boldsymbol\theta)$. This reduces reliance on the estimated cross-cell dependence \citep{schafer2005shrinkage,wickramasuriya2019optimal}. The matrix remains positive semidefinite throughout the parameter range, with $(\kappa,\chi)=(1,1)$ recovering $\mathbf V_P(\boldsymbol\theta)$ and $\kappa=0$ giving no added latent innovation.\footnote{In the numerical experiments and applications, we use a small set of centered Gaussian innovation distributions that vary in scale and in whether cross-cell dependence is retained, together with a specification without added latent innovation. We combine the corresponding predictive distributions using predictive stacking \citep{yao2018stacking,zhang2026geostatistics}. We describe the candidate distributions in Appendix~\ref{app:pcpa-innovation-library} and the stacking implementation in Appendix~\ref{app:pcpa-stacking}.}

To predict $\mathbf Y_{\mathcal M}(0)$, we combine the latent construction with the observation model. Under Assumption~\ref{ass:pcpa-conditional-observations}, the conditional observation law factorizes as
\[
p_{\mathcal M}
(d\mathbf y_{\mathcal M}\mid\mathbf z_{\mathcal M},\boldsymbol\theta)
=
\prod_{(i,t)\in\mathcal M}
p_{it}(dy_{it}\mid z_{it},\boldsymbol\theta).
\]
Let $\mathcal L_u(d\mathbf u_{\mathcal M}\mid \mathbf z_\Omega,\boldsymbol\theta)$ denote the joint conditional law of the latent innovations, satisfying the zero-mean condition in Equation~\eqref{eq:pcpa-latent-innovation}. For fixed $k$, integrating this law with the observation model gives the predictive distribution of $\mathbf Y_{\mathcal M}(0)$,
\begin{equation}
\begin{aligned}
\Pi^{\mathrm{pred}}_k
(d\mathbf y_{\mathcal M}\mid\mathcal D_\Omega)
={}&
\iint
\underbrace{
p_{\mathcal M}\!\left\{
d\mathbf y_{\mathcal M}\mid
c_k(\mathbf z_\Omega)+\mathbf u_{\mathcal M},
\boldsymbol\theta
\right\}
}_{\text{observation variation}}
\\
&\qquad\times
\underbrace{
\mathcal L_u
(d\mathbf u_{\mathcal M}\mid
\mathbf z_\Omega,\boldsymbol\theta)
}_{\text{latent variation}}
\underbrace{
\Pi_P
(d\mathbf z_\Omega,d\boldsymbol\theta
\mid\mathcal D_\Omega)
}_{\text{posterior uncertainty}} .
\end{aligned}
\label{eq:pcpa-predictive-law}
\end{equation}

Through the observation model, the innovation law affects the predictive dispersion and, for nonlinear links, the predictive mean.

For each joint predictive draw, we keep the observed treated outcomes fixed and compute
\begin{equation}
\tau_k^{(s)}(\mathbf q)
=\sum_{(i,t)\in\mathcal M}q_{it}
\left\{Y_{it}-Y_{k,it}^{(s)}(0)\right\}.
\label{eq:pcpa-predictive-effect}
\end{equation}
Aggregating within each joint draw retains dependence among the missing outcomes and carries it into uncertainty about $\tau(\mathbf q)$.

\subsection{Validation and tuning}
\label{subsec:pcpa-validation}
\label{subsec:model-selection}

In the causal panel data literature, tuning parameters, such as factor rank or regularization penalties, are typically selected by minimizing cross-validated squared prediction error on withheld untreated outcomes (e.g., pretreatment observations from treated units or observations from untreated units) \citep{xu2017generalized,athey2021matrix,ben2021augmented,liu2024practical}. The same approach can be used to select and tune causal estimators in our framework, as detailed in Appendix~\ref{app:pcpa-point-validation}. If several probabilistic models are considered, this selection can be repeated separately for each. The resulting minimized errors then provide a basis for comparing them. Because predictions are evaluated against the same withheld outcomes on the observed outcome scale, these comparisons remain meaningful even when the models have different latent scales or observation distributions. They can also include causal estimators applied directly to observed outcomes.\footnote{Models with similar point-prediction errors may imply different innovation distributions. In such cases, one could calibrate predictive distributions under each model and compare their predictive intervals for placebo effects using a proper interval score \citep{gneiting2007strictly}, at a common nominal level and with the weighting and aggregation of $\tau(\mathbf q)$.}

With the probabilistic model and causal estimator selected, predictive uncertainty can then be calibrated following the literature on forecast postprocessing \citep[e.g.,][]{sansom2016best}. Specifically, candidate innovation distributions or parameter values can be used to construct predictive distributions of placebo effects using the same weighting and aggregation as $\tau(\mathbf q)$, so that calibration targets the intended estimand. Because the withheld cells are untreated, their true effects are zero. We use predictive stacking \citep{yao2018stacking,zhang2026geostatistics} to combine the candidate predictive distributions, selecting the mixture weights by minimizing the average continuous ranked probability score (CRPS) at zero \citep{gneiting2007strictly}, as detailed in Appendix~\ref{app:pcpa-stacking}. How well these scores represent prediction for the treated aggregate depends on the validation design, particularly which cells are withheld. Under the convergence conditions stated in that appendix, the expected CRPS of the selected mixture approaches the minimum within the candidate family. For interval inference, convergence to the true conditional distribution gives asymptotically nominal coverage conditional on $\mathcal D_\Omega$ under the continuity conditions stated in Appendix~\ref{app:pcpa-stacking}.

Both selection and predictive calibration require posterior inference after validation outcomes are withheld. For latent Gaussian models fitted using \texttt{R-INLA} \citep{lindgren2015bayesian}, refitting for every fold can remain costly despite INLA's scalability \citep{rue2017bayesian}. Building on leave-group-out methods \citep{liu2025leavegroup,adin2024automatic}, we develop a fast cross-validation approximation that removes the withheld likelihood contributions at the fitted hyperparameter mode and samples from the resulting cavity approximation. Appendix~\ref{app:pcpa-cavity} details the approach and compares it with direct refitting. Posterior summaries agree closely in these comparisons, while posterior computation is more than a hundred times faster than refitting with hyperparameters re-estimated.

\section{Statistical properties}
\label{sec:pcpa-theory}
\label{sec:formal_analysis}

We study counterfactual imputation for a specified causal estimator $c_k$ and the posterior $\Pi_P(\cdot\mid\mathcal D_\Omega)$ defined by the probabilistic model. Separating uncertainty about the latent inputs from error in extrapolating to missing cells gives conditions for posterior consistency and a way to assess the consequences of pooling. We examine both the bias--variance trade-off for a fixed latent panel and an accuracy guarantee averaged over panels under correct probabilistic specification.

\subsection{Posterior consistency of counterfactual imputation}
\label{subsec:pcpa-consistency}
\label{subsec:imputation-learning}

Let $P_0$ denote the true distribution of the untreated outcome process, conditional on the realized treatment pattern and other fixed design information. Write $\mathbf z_\Omega^\star$ for the realized latent input panel and
\begin{equation}
\mathbf m_0(\mathbf z_\Omega^\star)
:=
\mathbb E_0\{\mathbf Z_{\mathcal M}(0)
\mid\mathbf Z_\Omega(0)=\mathbf z_\Omega^\star\}
\label{eq:pcpa-true-conditional-mean}
\end{equation}
\label{eq:true-conditional-mean}
for the true conditional mean of the missing latent outcomes.

Even with the true latent inputs, $c_k$ can make an imputation error. Uncertainty about those inputs introduces a further error whose magnitude depends on the estimator's sensitivity to perturbations. We measure input and imputation errors using the normalized Euclidean distances $d_\Omega(\mathbf z,\mathbf z')=|\Omega|^{-1/2} \|\mathbf z-\mathbf z'\|_2$ and $d_{\mathcal M}(\mathbf v,\mathbf v')=|\mathcal M|^{-1/2} \|\mathbf v-\mathbf v'\|_2$. Define
\begin{equation}
\begin{aligned}
b_k
&=d_{\mathcal M}\{c_k(\mathbf z_\Omega^\star),
                  \mathbf m_0(\mathbf z_\Omega^\star)\},\\
\zeta_k(r)
&=\sup_{d_\Omega(\mathbf z,\mathbf z_\Omega^\star)\leq r}
 d_{\mathcal M}\{c_k(\mathbf z),c_k(\mathbf z_\Omega^\star)\}.
\end{aligned}
\label{eq:pcpa-oracle-stability}
\end{equation}
The oracle error $b_k$ is the imputation error that $c_k$ would make if the true latent inputs were known. The function $\zeta_k(r)$ measures the largest change in its imputations when those inputs are perturbed within distance $r$.

For the asymptotic analysis, let $n$ index a sequence of panels in which the numbers of units and periods, or the precision of observations within cells, may vary. We use the subscript $n$ for quantities that vary along this sequence, with the normalized distances evaluated using the corresponding panel dimensions. The following assumption separates posterior concentration for the latent inputs from recovery and stability of the causal estimator.

\Needspace{8\baselineskip}
\begin{assumption}[Conditions for posterior consistency]
\label{ass:pcpa-imputation-consistency}
There is a deterministic sequence $r_n\downarrow0$ such that:
\begin{enumerate}
\renewcommand{\labelenumi}{(\roman{enumi})}
\item \emph{Posterior concentration.} The posterior for the latent inputs satisfies
\[
\Pi_P\!\left[
 d_\Omega\{\mathbf Z_{\Omega,n}(0),\mathbf z_{\Omega,n}^\star\}>r_n
 \mid\mathcal D_{\Omega,n}
\right]\longrightarrow0
\]
in $P_0$-probability.
\item \emph{Recovery at the true input.} $b_{k,n}\longrightarrow0$ in $P_0$-probability.
\item \emph{Local stability.} $\zeta_{k,n}(r_n)\longrightarrow0$ in $P_0$-probability.
\end{enumerate}
\end{assumption}

Whether part (i) holds depends on the observation model, prior, and asymptotic sequence. General posterior-contraction results provide tools for establishing this condition \citep{ghosal2007convergence}.\footnote{Increasing the number of cells alone need not remove uncertainty about each latent input. With one noisy measurement per cell, an independent latent component can retain nonvanishing posterior variance as the panel grows.} Parts (ii) and (iii) concern the causal estimator, through its identifying assumptions and its sensitivity to perturbations of the latent input, respectively.

For any latent input within distance $r$ of $\mathbf z_\Omega^\star$, the corresponding imputation is within $b_k+\zeta_k(r)$ of $\mathbf m_0(\mathbf z_\Omega^\star)$. This relationship yields a finite-sample posterior bound and, under Assumption~\ref{ass:pcpa-imputation-consistency}, posterior consistency.

\begin{theorem}[Transfer of posterior consistency]
\label{thm:pcpa-transfer}
\label{thm:imputation-transfer}
For every $r>0$,
\begin{equation}
\begin{aligned}
&\Pi_P\!\left[
 d_{\mathcal M}\{c_k(\mathbf Z_\Omega(0)),
                   \mathbf m_0(\mathbf z_\Omega^\star)\}
 > b_k+\zeta_k(r)\mid\mathcal D_\Omega
\right]\\
&\qquad\leq
\Pi_P\!\left[
 d_\Omega\{\mathbf Z_\Omega(0),\mathbf z_\Omega^\star\}>r
 \mid\mathcal D_\Omega
\right].
\end{aligned}
\label{eq:pcpa-transfer-bound}
\end{equation}
\label{eq:imputation-transfer-bound}
Under Assumption~\ref{ass:pcpa-imputation-consistency}, the imputation posterior concentrates on the true conditional mean $\mathbf m_{0,n}(\mathbf z_{\Omega,n}^\star)$ in $P_0$-probability. It contracts at rate $r_n$ if $b_{k,n}=O_{P_0}(r_n)$ and $c_{k,n}$ is Lipschitz on the $r_n$-neighborhood of the true input with constant $L_{k,n}=O_{P_0}(1)$.
\end{theorem}

\begin{proof}
See Appendix~\ref{app:pcpa-transfer-details}.
\end{proof}

The theorem shows that accurate learning of the latent inputs carries over to counterfactual imputation when the causal estimator can recover the missing conditional mean and remains stable under input error. Under the additional rate conditions, the imputation posterior concentrates at the same rate as the posterior for the latent inputs.

The recovery and stability conditions depend on the causal estimator. For synthetic controls, suppose that $\mathbf z_\Omega^\star$ and $\mathbf m_0(\mathbf z_\Omega^\star)$ share an additive factor representation. If a convex combination of controls reproduces the treated unit's intercept and factor loadings, and the pretreatment factor design, including an intercept, has full column rank, exact pretreatment matching gives $b_k=0$ \citep{abadie2021using}. Perturbations of the latent input affect both the fitted control weights and the control outcomes used for imputation. When the weight solution is unique and the objective satisfies local quadratic growth, optimization perturbation results provide bounds on changes in the weights \citep{bonnans2000perturbation}. Combined with the direct changes in control outcomes, these bounds control $\zeta_k(r)$. We discuss the corresponding recovery and stability conditions for interactive fixed effects and matrix completion in Appendix~\ref{app:pcpa-recovery}.

\subsection{When does pooling improve counterfactual imputation?}
\label{subsec:pcpa-pooling}
\label{subsec:pooling-risk}

Existing analyses characterize how noise and regularization affect imputation error within particular causal panel estimators \citep{amjad2018robust,athey2021matrix,ben2021augmented,ferman2021synthetic}. In our framework, pooling takes place before and separately from the causal estimator. We show that its effect on imputation accuracy is governed by a bias--variance tradeoff in which changes in variance and smoothing bias are weighted according to how perturbations of the latent input propagate to the imputed counterfactual.

To compare imputation risk with and without pooling, suppose that $c_k$ is affine in its latent input,
\begin{equation}
c_k(\mathbf z)=c_k(\mathbf z_\Omega^\star)
+\mathbf J_k(\mathbf z-\mathbf z_\Omega^\star),
\label{eq:pcpa-affine-map}
\end{equation}
\label{eq:affine-imputation-map}
where $\mathbf J_k\in\mathbb R^{|\mathcal M|\times|\Omega|}$ describes the sensitivity of the imputation to perturbations of the input. For differentiable estimators, the corresponding first-order risk calculation uses the Jacobian at $\mathbf z_\Omega^\star$, provided the contribution of the higher-order remainder is negligible.

To isolate the effect of pooling itself, we set \(b_k=0\), equivalently \(c_k(\vect z_\Omega^\star)=\vect m_0(\vect z_\Omega^\star)\), and condition on the realized latent input \(\vect z_\Omega^\star\). Let \(\widehat{\vect z}_U=\vect z_\Omega^\star+\vect\varepsilon\) denote an unpooled estimate and \(\widehat{\vect z}_P=\mat S\widehat{\vect z}_U\) its pooled counterpart, where \(\mat S\) is a fixed linear pooling operator. We assume \(\E_0[\vect\varepsilon\mid \vect Z_\Omega(0)=\vect z_\Omega^\star]=\vect 0\) and \(\Var_0[\vect\varepsilon\mid \vect Z_\Omega(0)=\vect z_\Omega^\star]=\mat\Sigma\).

For an estimator \(\widehat{\vect z}\) of the latent input panel, define its conditional imputation risk by
\begin{equation}
	\mathcal R_k
	\left(
	\widehat{\vect z}
	\mid
	\vect z_\Omega^\star
	\right)
	=
	\E_0
	\left[
	d_{\mathcal M}^2
	\left\{
	c_k(\widehat{\vect z}),
	\vect m_0(\vect z_\Omega^\star)
	\right\}
	\mid
	\vect Z_\Omega(0)=\vect z_\Omega^\star
	\right].
	\label{eq:pcpa-imputation-risk}
\end{equation}
\label{eq:imputation-risk}

Write \(\mat H_k=\mat J_k\trans\mat J_k\) and let \(\|\vect v\|_{\mat H_k}^2=\vect v\trans\mat H_k\vect v\) denote the corresponding quadratic form.

The following result expresses the change in imputation risk as squared smoothing bias plus the change in variance, both evaluated after propagation through the causal estimator.

\begin{proposition}[Pooling and imputation risk]
\label{prop:pcpa-pooling}
\label{prop:pooling-risk}
Under the affine representation in Equation~\eqref{eq:pcpa-affine-map}, $b_k=0$, a fixed pooling operator $\mathbf S$, and the conditional error moments stated above,
	\begin{align}
		|\mathcal M|
		\left\{
		\mathcal R_k
		\left(
		\widehat{\vect z}_P
		\mid
		\vect z_\Omega^\star
		\right)
		-
		\mathcal R_k
		\left(
		\widehat{\vect z}_U
		\mid
		\vect z_\Omega^\star
		\right)
		\right\}
		={}&
		\left\|
		(\mat S-\mat I)
		\vect z_\Omega^\star
		\right\|_{\mat H_k}^2
		\nonumber\\
		&-
		\operatorname{tr}
		\left[
		\mat H_k
		\left\{
		\mat\Sigma
		-
		\mat S\mat\Sigma\mat S\trans
		\right\}
		\right].
		\label{eq:pcpa-pooling-identity}
	\end{align}
\label{eq:pooling-risk-difference}
\end{proposition}

\begin{proof}
See Appendix~\ref{app:pcpa-pooling-proof}.
\end{proof}

Proposition~\ref{prop:pcpa-pooling} shows that pooling lowers imputation risk when its reduction in the variance of the counterfactual estimates exceeds the squared bias it introduces. The matrix $\mathbf H_k$ weights input errors according to their effect on counterfactual imputation. For example, in Equation~\eqref{eq:pcpa-affine-map}, a perturbation $\mathbf v$ satisfying $\mathbf J_k\mathbf v=0$ leaves the imputation unchanged. Better cellwise reconstruction can therefore coexist with higher imputation risk if pooling distorts features of the latent panel that matter for the causal estimator. This estimator dependence motivates the counterfactual prediction criterion for model selection in Section~\ref{subsec:pcpa-validation}, rather than criteria that evaluate the hierarchical model independently of the causal estimator.

Pooling can also change the factor or low-rank structure recovered by interactive fixed-effects and matrix completion estimators. When the relevant signal directions are preserved, smoothing can improve factor recovery \citep{fan2016projected}. We illustrate this mechanism for a Gaussian space--time model in Appendix~\ref{app:pcpa-low-rank-details}.

Proposition~\ref{prop:pcpa-pooling} compares imputation risk for a fixed latent panel. Under a correctly specified posterior for the latent inputs, we also derive an accuracy guarantee for estimates obtained by averaging imputations across posterior draws. The result applies to general causal estimators, including nonlinear ones. Write
\[
\bar{\mathbf c}_k(\mathcal D_\Omega)
=
\mathbb E_P\!\left[
 c_k\{\mathbf Z_\Omega(0)\}\mid\mathcal D_\Omega
\right]
\]
for the posterior expectation of the imputation, approximated by averaging the draws in Equation~\eqref{eq:pcpa-draw-wise}. For a measurable predictor $\mathbf a$ based on $\mathcal D_\Omega$, define its Bayes risk, or expected squared imputation error, by
\[
\mathcal B(\mathbf a)
=
\mathbb E_0\!\left[
 d_{\mathcal M}^2\left\{
  \mathbf a(\mathcal D_\Omega),
  \mathbf m_0\{\mathbf Z_\Omega(0)\}
 \right\}
\right].
\]
The expectation is taken over the joint distribution of $\mathbf Z_\Omega(0)$ and $\mathcal D_\Omega$ under $P_0$. Taking the infimum over all such predictors gives $\inf_{\mathbf a}\mathcal B(\mathbf a)$, the smallest expected squared error attainable for the missing latent conditional mean using these data.

\begin{proposition}[Bayes risk of counterfactual imputation]
\label{prop:pcpa-bayes-risk}
Suppose that the conditional distribution of $\mathbf Z_\Omega(0)$ given $\mathcal D_\Omega$ under the probabilistic model equals that under $P_0$, $P_0$-almost surely, and that $c_k\{\mathbf Z_\Omega(0)\}$ and $\mathbf m_0\{\mathbf Z_\Omega(0)\}$ have finite second moments under $P_0$. Then
\[
0\leq
\mathcal B(\bar{\mathbf c}_k)
-\inf_{\mathbf a}\mathcal B(\mathbf a)
\leq
\mathbb E_0\!\left[
 d_{\mathcal M}^2\left\{
  c_k\{\mathbf Z_\Omega(0)\},
  \mathbf m_0\{\mathbf Z_\Omega(0)\}
 \right\}
\right].
\]
\end{proposition}

\begin{proof}
See Appendix~\ref{app:pcpa-pooling-proof}.
\end{proof}

Proposition~\ref{prop:pcpa-bayes-risk} shows that, under a correctly specified posterior for the latent inputs, the excess Bayes risk of our estimator is bounded by the expected imputation error that $c_k$ would make even if those inputs were known. If $c_k$ recovers the missing latent conditional mean from the true inputs, our estimator therefore achieves the lowest possible expected squared imputation error among all estimators based on $\mathcal D_\Omega$, including causal panel estimators applied directly to observed outcomes.

\section{Numerical evidence and empirical illustrations}
\label{sec:pcpa-evidence}

The theoretical results give conditions for posterior consistency of counterfactual imputation and describe the trade-off between smoothing bias and variance reduction. We now examine this trade-off in finite samples and assess predictive inference for realized treatment effects. We then compare probabilistic specifications in three policy evaluations. Full experimental designs, application details, and additional results are reported in Appendices~\ref{app:pcpa-designs} and~\ref{app:pcpa-empirical}.

\subsection{Numerical evidence}
\label{subsec:pcpa-numerical-evidence}
\label{sec:pcpa-experiments}
\label{sec:num_experiments}

The benefit of pooling for factor-based imputation depends on whether it preserves the signal from which the factor structure is recovered. Spatial smoothing can reduce observation noise that obscures common factors, but may also attenuate genuine variation in factor loadings when the signal is not spatially structured. To examine this trade-off, we apply generalized synthetic controls (GSC; \citealp{xu2017generalized}) to Gaussian panels with varying observation noise and spatial structure in the factor loadings. We compare estimation directly from observed outcomes with estimation based on Gaussian probabilistic models with and without spatial pooling (Appendix~\ref{app:experiment1-design}).\footnote{Both models include unit-by-period interactions with AR(1) temporal dependence. Across units, the nonspatial interaction is independent, while the Besag--York--Molli\'e (BYM) specification \citep{besag1991bayesian} combines geographically structured and unstructured components. For computational efficiency, this experiment uses separate component precisions; Appendix~\ref{app:experiment1-design} explains the distinction from the BYM2 parameterization used in the other spatial analyses.}

\begin{figure}[!htbp]
\centering
\caption{Pooling and counterfactual imputation accuracy.}
\includegraphics[width=\textwidth]{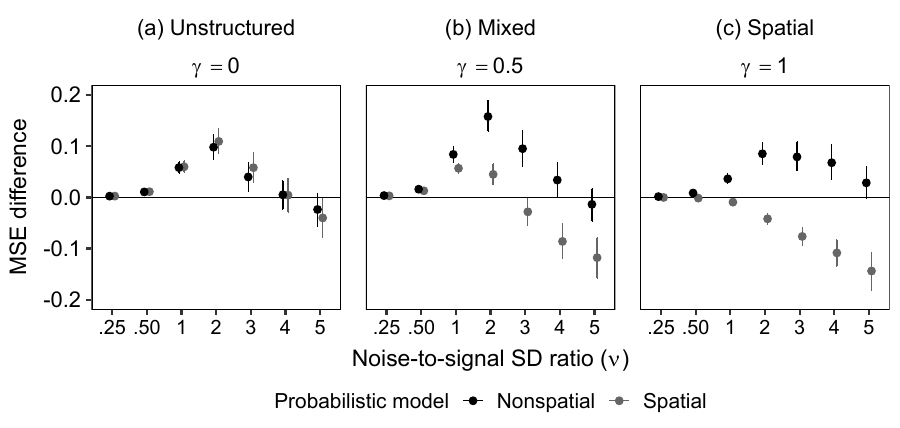}
\figurenote{The vertical axis shows the difference in mean squared error (MSE) for the average counterfactual over 100 treated cells relative to GSC fitted directly to observed outcomes; negative values favor the probabilistic analysis. Panels (a)--(c) vary the spatially structured share of the factor loadings ($\gamma=0,0.5,1$), while $\nu$ on the horizontal axis is the observation-error standard deviation (approximately the noise-to-signal standard deviation ratio). Black points use a Gaussian model with interactions that are independent across units, and gray points use Besag--York--Molli\'{e} (BYM) spatial pooling. Both models have AR(1) temporal dependence. Simulations use 60 units, 30 periods, ten units treated after period 20, and factor rank two. Error bars give pointwise 95\% Monte Carlo intervals from paired bootstrap resampling of 1,000 replications.}
\label{fig:pcpa-risk}
\label{fig:experiment-center}
\end{figure}

Figure~\ref{fig:pcpa-risk} illustrates how the benefits of smoothing depend on the amount of observation noise and the structure available for pooling. The largest gains occur at higher noise-to-signal standard deviation ratios, particularly when the graph structure used by the probabilistic model aligns with the spatial structure of the factor loadings. When there is limited structure to exploit, smoothing can instead increase counterfactual imputation error.

In a second Gaussian experiment, we examine predictive inference for average treatment effects when latent variation remains around the imputed counterfactuals. We vary its magnitude and dependence and compare calibrated prediction with prediction that omits this latent variation. Both include posterior uncertainty and observation variation, and use the same fitted probabilistic model and GSC rank selected by cross-validation within each replication (Appendix~\ref{app:experiment2-design}).

The calibrated mixture brings interval coverage closer to the nominal 95\% level, with the largest improvement under substantial correlated latent variation. Predictive scores improve in that setting and change little elsewhere (Table~\ref{tab:pcpa-predictive-performance}).

\begin{table}[!htbp]
\centering
\begin{threeparttable}
\caption{Coverage and CRPS differences for the sample average~treatment~effect~on~the~treated.}
\label{tab:pcpa-predictive-performance}
\small
\setlength{\tabcolsep}{4pt}
\begin{tabular*}{\textwidth}{@{\extracolsep{\fill}}l
 S[table-format=2.1] S[table-format=2.1]
 S[table-format=-1.1] c@{}}
\toprule
 & \multicolumn{2}{c}{Coverage (\%)}
 & \multicolumn{2}{c}{CRPS difference ($\times 10^3$)}\\
\cmidrule(lr){2-3}\cmidrule(lr){4-5}
Simulated innovations
 & {\shortstack[c]{No added latent\\innovations}}
 & {\shortstack[c]{Calibrated\\prediction}}
 & {Estimate}
 & \shortstack[c]{95\% Monte Carlo\\interval}\\
\midrule
None                       & 92.8 & 94.8 & -0.5 & $[-1.3,\phantom{-}0.3]$\\
Independent                & 91.8 & 94.4 & -0.5 & $[-1.3,\phantom{-}0.4]$\\
Mixed                      & 92.0 & 94.2 & -0.5 & $[-1.4,\phantom{-}0.4]$\\
Correlated                 & 90.4 & 93.2 & -0.6 & $[-1.5,\phantom{-}0.3]$\\
Correlated, doubled scale  & 88.4 & 93.8 & -2.1 & $[-3.7,-0.6]$\\
\bottomrule
\end{tabular*}
\begin{tablenotes}[flushleft]
\footnotesize
\item Notes: Across 500 replications per regime, the table reports coverage of 95\% predictive intervals for a realized average treatment effect of 0.50 over 100 treated cells. The simulated latent innovations range from none to independent, mixed, and correlated variation; the final regime doubles the innovation standard deviation. Both predictions use the same spatial Gaussian model and cross-validated GSC rank and account for posterior uncertainty and observation variation. Calibrated prediction also allows latent innovations whose scale and dependence are calibrated using held-out untreated outcomes. The continuous ranked probability score (CRPS) difference is calibrated prediction minus prediction without added latent innovations; negative values favor calibration. The final column gives paired 95\% Monte Carlo intervals for these score differences.
\end{tablenotes}
\end{threeparttable}
\end{table}

In Appendix~\ref{app:experiment3-design}, we also examine the framework with simulated Poisson counts using synthetic controls (SCM) and ridge-augmented synthetic controls (ASCM; \citealp{ben2021augmented}). Comparisons with empirical log rates distinguish the contribution of probabilistic modeling from that of changing the estimation scale. Both estimators have lower imputation error with the Poisson model than with empirical log rates. Coverage of the $95\%$ predictive intervals ranges from $93\%$ to $98\%$ across the examined settings (Table~\ref{tab:pcpa-poisson}).

\subsection{Empirical illustrations}
\label{subsec:pcpa-empirical-illustrations}
\label{sec:pcpa-applications}
\label{sec:applications}

The potential gains from probabilistic modeling depend on the information contained in the observed outcomes and the structure available for counterfactual imputation. We examine these gains in three established policy evaluations with different measurement and estimation challenges. Table~\ref{tab:pcpa-empirical-summary} summarizes the results, while Appendix~\ref{app:pcpa-applications} provides the data, model specifications, validation designs, and additional analyses.

\begin{table}[!htbp]
\centering
\begin{threeparttable}
\caption{Treatment-effect estimates, intervals, and cross-validated RMSE in~three~policy~evaluations.}
\label{tab:pcpa-empirical-summary}
\label{tab:pcpa-california}
\label{tab:pcpa-cannabis}
\label{tab:pcpa-vision-zero}
\small
\setlength{\tabcolsep}{4pt}
\begin{tabular*}{\textwidth}{@{\extracolsep{\fill}}ll
 S[table-format=-2.2] c S[table-format=2.2]@{}}
\toprule
Method & \shortstack[l]{Data or\\probabilistic model}
 & {\shortstack[c]{Average\\effect}}
 & 95\% interval & {\shortstack[c]{Cross-validated\\RMSE}}\\
\midrule
\multicolumn{5}{@{}l}{California's tobacco-control program}\\
SCM & Observed sales       & -18.8 & \textemdash    & 12.44\\
SCM & Gaussian, nonspatial & -18.8 & $[-42.5,5.3]$  & 13.20\\
SCM & Gaussian, spatial    & -18.7 & $[-40.7,3.9]$  & 12.24\\
\addlinespace[0.6em]
\multicolumn{5}{@{}l}{Recreational cannabis laws and opioid mortality}\\
GSC & Observed rates       & -0.25 & $[-2.79,2.30]$ & 1.91\\
GSC & Poisson, nonspatial  & -0.95 & $[-4.20,1.04]$ & 1.88\\
GSC & Poisson, spatial     & -0.62 & $[-3.53,1.36]$ & 1.66\\
\addlinespace[0.6em]
\multicolumn{5}{@{}l}{Vision Zero and traffic fatalities}\\
SCM  & Observed rates      & -1.39 & \textemdash    & 2.88\\
ASCM & Observed rates      & 0.15  & \textemdash    & 0.91\\
ASCM & Poisson, nonspatial & 0.25  & $[-3.56,1.96]$ & 0.52\\
\bottomrule
\end{tabular*}
\begin{tablenotes}[flushleft]
\footnotesize
\item Notes: SCM, GSC, and ASCM denote synthetic controls, generalized synthetic controls, and ridge-augmented synthetic controls. Effects are averages over 1989--2000 for tobacco control (packs per capita), twenty treated state-years for cannabis legalization (deaths per 100,000), and 1997--2015 for Vision Zero (fatalities per billion passenger-kilometers). For the probabilistic analyses, we report predictive medians and 95\% predictive intervals. The interval for GSC fitted to observed opioid-mortality rates is instead a parametric cluster-bootstrap confidence interval \citep{xu2017generalized}. Dashes mean that no aggregate interval is reported. Cross-validated RMSE measures error on withheld untreated outcomes in the same units as the corresponding effect; smaller values indicate better prediction within an application. Model and tuning details are given in Appendix~\ref{app:pcpa-empirical}.
\end{tablenotes}
\end{threeparttable}
\end{table}

\FloatBarrier

We begin with California's Proposition 99 tobacco-control program, a widely studied application of synthetic controls \citep{abadie2010synthetic}. The original analysis estimates how cigarette sales would have evolved in California without the program, using other states to construct the counterfactual. Annual cigarette sales over a long pretreatment period provide comparatively informative data for matching, leaving relatively little scope for probabilistic modeling to improve the inputs to SCM. Consistent with this expectation, the estimated counterfactual trajectory and average treatment effect change little when SCM is applied to posterior draws rather than observed sales. The spatial Gaussian model improves cross-validated prediction only marginally. This is a setting where the additional modeling has limited practical consequences, as the observed outcomes already provide a relatively informative basis for counterfactual estimation.

\begin{figure}[!htbp]
\centering
\caption{Recreational cannabis laws and opioid mortality.}
\includegraphics[width=\textwidth]{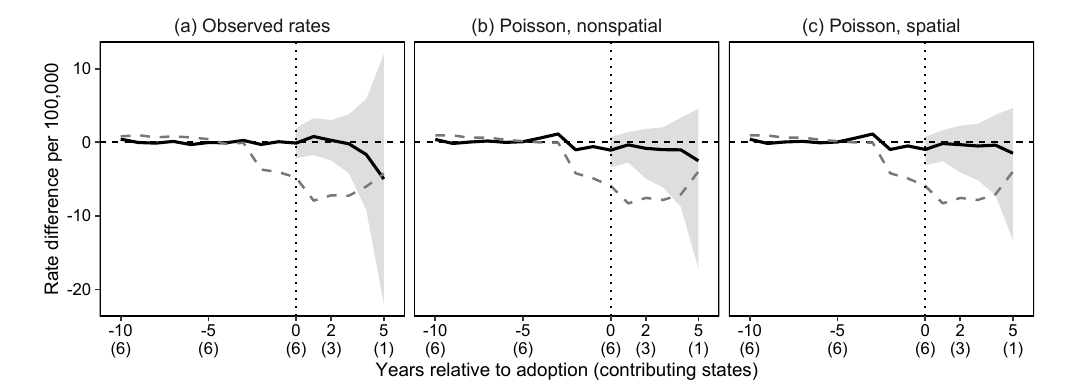}
\figurenote{Event-time estimates of observed minus counterfactual opioid-mortality rates (deaths per 100,000), averaged equally over contributing states. Solid curves show GSC estimates using (a) observed rates, (b) posterior draws from a nonspatial Poisson model, and (c) posterior draws from a spatial Poisson model. Gray dashed curves give additive-model imputations from a linear model on observed rates in (a) and the same additive Poisson fit in (b)--(c). The additive models adjust for four policy covariates and unit and period effects but have no unit-by-period interactions. Shaded bands apply only to GSC: pointwise 95\% residual-bootstrap confidence intervals in (a) and posterior predictive intervals in (b)--(c). Numbers below the axis give the number of contributing states at each event time. The panel covers 46 U.S. jurisdictions in 1999--2019, including six staggered adopters.}
\label{fig:pcpa-cannabis}
\label{fig:pcpa-cannabis-event-study}
\label{fig:marijuana}
\end{figure}

Recreational cannabis laws and opioid mortality present a different problem. We revisit the evaluation of staggered cannabis legalization across U.S. states \citep{nguyen2024cannabis}, where mortality is measured through death counts with varying population exposures. The precision of observed mortality rates differs across states and years, and geographic dependence may provide useful information about underlying mortality trajectories. Here, fitting GSC to latent log rates from a nonspatial Poisson model improves cross-validated prediction only slightly relative to GSC on observed rates. Adding spatial pooling produces a more substantial improvement, suggesting that geographic structure helps recover information relevant to counterfactual prediction. The estimated effects of legalization remain broadly similar across specifications, although the spatial model gives somewhat narrower predictive intervals. The main gain in this application is therefore more accurate prediction of withheld untreated outcomes rather than a substantial change in the estimated policy effect (Figure~\ref{fig:pcpa-cannabis-event-study}).

\FloatBarrier

\begin{figure}[!htbp]
\centering
\caption{Vision Zero and traffic fatalities.}
\includegraphics[width=\textwidth]{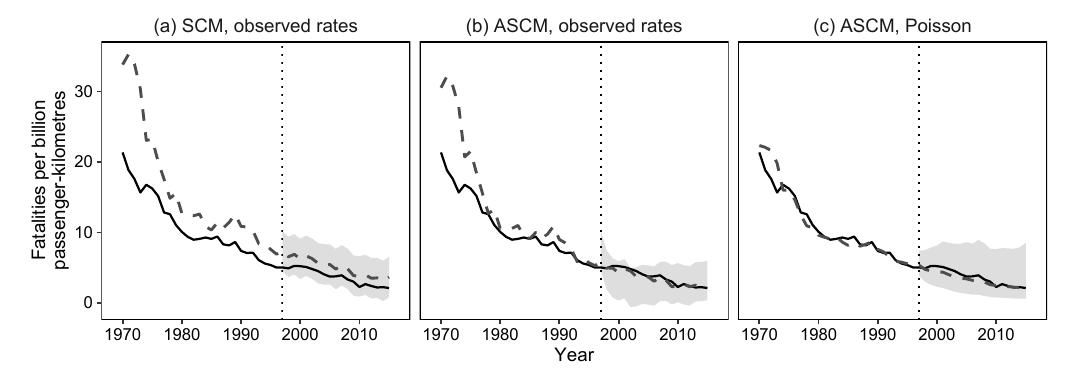}
\figurenote{Sweden's observed fatality rates (solid) and estimated counterfactuals (dashed) using (a) SCM on observed rates, (b) ASCM on observed rates, and (c) ASCM with a Poisson observation model. The panel includes Sweden and nine controls in 1970--2015; vertical dotted lines mark adoption in 1997. Rates are fatalities per billion passenger-kilometers. Shaded bands are 95\% pointwise conformal intervals in (a)--(b) \citep{chernozhukov2021conformal} and posterior predictive intervals in (c).}
\label{fig:pcpa-vision-zero}
\label{fig:vision-zero}
\end{figure}

Our final example revisits Sweden's Vision Zero road-safety policy, introduced in 1997 \citep{bonander2021flexible}. Sweden's low pretreatment traffic-fatality rates cannot be reproduced as a nonnegative weighted average of the control-country rates. Ridge augmentation permits extrapolation beyond the donor convex hull and substantially improves prediction \citep{ben2021augmented}, but may also lead to negative counterfactuals. Combining the augmented estimator with a Poisson model of traffic-fatality counts improves prediction further, while keeping extrapolated counterfactual mean rates positive (Figure~\ref{fig:pcpa-vision-zero}).

\FloatBarrier

\section{Discussion}
\label{sec:pcpa-discussion}
\label{sec:discussion}

Our framework allows researchers to combine established causal panel estimators with observation models and structured pooling suited to their data, without developing a separate joint Bayesian formulation for each estimator. Implementation is straightforward with existing causal estimation routines and any suitable probabilistic model that supplies joint posterior draws of the latent panel and its parameters.

The analyses here begin with a unit-period panel constructed in advance. In many applications, however, panel outcomes are themselves estimated from lower-level observations. The probabilistic model could operate directly on these data, allowing uncertainty in constructing the panel to propagate through counterfactual imputation. In repeated cross-sectional surveys, for example, researchers could model individual responses and standardize posterior draws to a common target population, accounting for unequal sample sizes, survey weights, and changing covariate composition \citep{little1993poststratification,si2020bayesian}. Similarly, observations recorded at finer spatial or temporal resolution could be modeled on their native support \citep{lindgren2011explicit}, with posterior draws aggregated to the units and periods required by the causal design.

Our theoretical analysis concerns the general construction. Further work on particular combinations of smoothers and causal estimators could establish sharper conditions under which pooling improves counterfactual imputation. For predictive inference, a related goal is to derive the limiting frequentist coverage of intervals for aggregate treatment effects. The value of such results is illustrated by \citet{wang2023coverage}, who characterize the limiting coverage of credible intervals from transformed posteriors in multivariate isotonic regression and use it to recalibrate credibility levels. Extending this type of analysis to causal panel estimators would require accounting for imputation error and the dependence of the remaining outcome variation, establishing when predictive intervals achieve nominal coverage and whether further calibration is needed.

In practice, model comparisons should assess counterfactual predictions and include the same causal estimator fitted directly to observed outcomes. Our fast cross-validation procedure makes these comparisons and predictive calibration practical for latent Gaussian models. Their usefulness nevertheless depends on the adequacy of the probabilistic model and on how well the validation problems represent prediction for the treated cells. Further work could draw on Bayesian inference under misspecification and generalized updating \citep{muller2013risk,bissiri2016general} to improve the robustness of probabilistic modeling, and on methods for prediction under covariate shift \citep{shimodaira2000improving,sugiyama2007covariate} to adapt validation when the available untreated observations differ systematically from the treated cells.

\clearpage
\bibliographystyle{apalike}
\bibliography{references}

\newgeometry{margin=20mm,footskip=29bp}
\appendix
\setcounter{figure}{0}
\setcounter{table}{0}
\renewcommand{\thefigure}{A.\arabic{figure}}
\renewcommand{\thetable}{A.\arabic{table}}

\addtocontents{toc}{\protect\setcounter{tocdepth}{1}}
\addtocontents{toc}{\protect\setcounter{tocdepth}{1}}
\addtocontents{toc}{\protect\setcounter{tocdepth}{1}}
\addtocontents{toc}{\protect\setcounter{tocdepth}{1}}
\vspace*{-0.8em}
\begin{center}
{\Large\bfseries Online Appendix\par}
  \vspace{1.7em}
{\LARGE Probabilistic Causal Panel Analysis\par}
  \vspace{1.8em}
{\large Carl Bonander, Qinyun Lin, Ulf Str\"omberg and Marta Blangiardo\par}
\end{center}

\vspace{2.5em}

\noindent This supplement includes implementation details, proofs, and other supporting information. The numerical experiments and policy evaluations summarized in Section~\ref{sec:pcpa-evidence} are reported with their full designs and results.

\vspace{1.5em}
\begingroup
\renewcommand{\contentsname}{Contents}
\tableofcontents
\endgroup
\clearpage

\section{Cross-validation and predictive inference}
\label{app:pcpa-predictive}
\label{app:predictive-construction}

This appendix describes the validation criteria and predictive calculations used in Section~\ref{subsec:pcpa-validation}. We compare probabilistic models and causal estimators through their predictions of held-out untreated outcomes. To calibrate the predictive distribution, we then assess uncertainty after aggregating predictions in the same way as the treatment effect of interest.

\subsection{Validation design and point prediction}
\label{app:pcpa-point-validation}

We choose validation blocks to reflect the information available for counterfactual prediction. For example, when a block begins at a hypothetical intervention date, later observations from the same unit may also need to be withheld, even if only part of that period is used to assess prediction. For fold $\ell$, let $D_\ell\subseteq\Omega$ contain all cells whose observations are removed and let $V_\ell\subseteq D_\ell$ contain the cells used to calculate the prediction error. The posterior is conditional on $\mathcal D_{\Omega\setminus D_\ell}$. Observations in $\mathcal M$ remain excluded, and a unit used for validation is excluded from its own control pool.

For a specified probabilistic model and causal estimator $k$, we apply the construction in Section~\ref{subsec:pcpa-draw-wise} with $D_\ell$ treated as missing. The notation $c_k$ refers to the same estimator and tuning parameters, now applied under this reduced observation pattern. The superscript $(s,-\ell)$ denotes joint posterior draw $s$ after withholding $D_\ell$. For $s=1,\ldots,S$, define
\begin{equation}
C_{k,it}^{(s,-\ell)}
=c_{k,it}\{\mathbf Z_{\Omega\setminus D_\ell}^{(s,-\ell)}(0)\},
\qquad (i,t)\in V_\ell.
\label{eq:pcpa-fold-imputation}
\end{equation}
For point prediction, we map each imputation through the observation mean and average over posterior draws,
\begin{equation}
\begin{aligned}
\mu_{k,it}^{(s,-\ell)}
&=\int y\,p_{it}\{dy\mid C_{k,it}^{(s,-\ell)},
                         \boldsymbol\theta^{(s,-\ell)}\},\\
\overline\mu_{k,it}^{(-\ell)}
&=S^{-1}\sum_{s=1}^S\mu_{k,it}^{(s,-\ell)}.
\end{aligned}
\label{eq:pcpa-point-means}
\end{equation}
\label{eq:center-validation-mean}
For a Poisson model with exposure $E_{it}$ and log-rate imputation, $\mu_{k,it}^{(s,-\ell)}=E_{it}\exp\{C_{k,it}^{(s,-\ell)}\}$. Exponentiation precedes averaging over posterior draws, so uncertainty on the latent scale contributes to the point prediction on the outcome scale.

For a specified probabilistic model, we select the causal estimator and its tuning parameters by minimizing average squared prediction error,
\begin{equation}
L_{\mathrm{point}}(k)
=\frac1L\sum_{\ell=1}^L\frac1{|V_\ell|}
 \sum_{(i,t)\in V_\ell}
 \{Y_{it}-\overline\mu_{k,it}^{(-\ell)}\}^2.
\label{eq:pcpa-point-score}
\end{equation}
\label{eq:point-validation-loss}
This expression is on the observation scale. For rate comparisons, we divide both observations and predictions by exposure and express them in the same units before calculating the squared error. The transformation is the same for all specifications being compared. If several probabilistic models are considered, we minimize the criterion separately for each and compare the attained errors. For each subsequent predictive analysis, the probabilistic model and selected causal estimator, including its tuning parameters, are held fixed when estimating the mixture weights.

\subsection{Candidate innovation distributions}
\label{app:pcpa-innovation-library}

For the numerical experiments and empirical analyses, we combine centered Gaussian innovation distributions that differ in scale and cross-cell dependence. Their covariances are based on $\mathbf D_P(\boldsymbol\theta)$ and $\mathbf V_P(\boldsymbol\theta)$ from Equation~\eqref{eq:pcpa-reference-covariance}.

For each posterior draw, the candidates share the counterfactual imputation $\widehat{\mathbf Z}_{k,\mathcal M}^{(s)}(0)$ from Equation~\eqref{eq:pcpa-draw-wise}. Conditional on $\boldsymbol\theta^{(s)}$, we generate the two reference innovations from
\[
\mathbf e_{I,\mathcal M}^{(s)}
\sim\mathcal N\{\mathbf0,\mathbf D_P(\boldsymbol\theta^{(s)})\},
\qquad
\mathbf e_{J,\mathcal M}^{(s)}
\sim\mathcal N\{\mathbf0,\mathbf V_P(\boldsymbol\theta^{(s)})\},
\]
where $I$ denotes innovations independent across missing cells and $J$ preserves the model's estimated conditional dependence. Combining these with scale multipliers $\kappa\in\{1,2\}$ and a specification without added latent innovations gives
\begin{equation}
\begin{aligned}
\mathbf Z_{0,k,\mathcal M}^{(s)}(0)
 &=\widehat{\mathbf Z}_{k,\mathcal M}^{(s)}(0),\\
\mathbf Z_{I,\kappa,k,\mathcal M}^{(s)}(0)
 &=\widehat{\mathbf Z}_{k,\mathcal M}^{(s)}(0)
   +\kappa\mathbf e_{I,\mathcal M}^{(s)},\\
\mathbf Z_{J,\kappa,k,\mathcal M}^{(s)}(0)
 &=\widehat{\mathbf Z}_{k,\mathcal M}^{(s)}(0)
   +\kappa\mathbf e_{J,\mathcal M}^{(s)},
   \qquad \kappa\in\{1,2\}.
\end{aligned}
\label{eq:pcpa-candidate-fields}
\end{equation}
\label{eq:predictive-library-components}

Write $\mathcal J=\{0,I1,J1,I2,J2\}$ for the resulting candidate set. For each candidate, we generate untreated outcomes through the observation model and form the treatment-effect contrast in Equation~\eqref{eq:pcpa-predictive-effect} within each joint draw. Let $Q_{j,k}$ denote the resulting predictive distribution of $\tau(\mathbf q)$.

Although the candidate scales are discrete, their mixture permits continuous adjustment of the added latent covariance. Because the candidates have the same conditional latent mean, mixture weights $w_j\geq0$ summing to one give
\begin{equation}
\mathbf K_{\mathbf w}(\boldsymbol\theta)
=(w_{I1}+4w_{I2})\mathbf D_P(\boldsymbol\theta)
 +(w_{J1}+4w_{J2})\mathbf V_P(\boldsymbol\theta).
\label{eq:pcpa-mixture-covariance}
\end{equation}
\label{eq:stacked-reference-covariance}

With $a=w_{I1}+4w_{I2}$ and $b=w_{J1}+4w_{J2}$, this covariance corresponds to $\kappa=\sqrt{a+b}$ and $\chi=b/(a+b)$ in Equation~\eqref{eq:pcpa-covariance-family} when $a+b>0$; at $a+b=0$ there is no added innovation. Varying the weights covers $0\leq\kappa\leq2$ and $0\leq\chi\leq1$, including scales below one through weight on the no-innovation candidate. This equivalence concerns the conditional covariance; the full predictive distribution remains a mixture and need not be Gaussian. We examine larger candidate scales in Appendix~\ref{app:pcpa-empirical-sensitivity}.

\subsection{Predictive calibration by stacking}
\label{app:pcpa-stacking}

We combine the candidate predictive distributions using predictive stacking \citep{yao2018stacking,zhang2026geostatistics},
\begin{equation}
Q_{\mathbf w,k}=\sum_{j\in\mathcal J}w_jQ_{j,k},
\qquad
\mathbf w\in\Delta_{\mathcal J}
=\{\mathbf w:w_j\geq0,\ \textstyle\sum_{j\in\mathcal J}w_j=1\}.
\label{eq:pcpa-predictive-mixture}
\end{equation}
\label{eq:stacked-predictive-distribution}
Because the candidates share the posterior, causal estimator, and observation model, this mixture is obtained from Equation~\eqref{eq:pcpa-predictive-law} by using the innovation law $\mathcal L_{u,\mathbf w}=\sum_{j\in\mathcal J}w_j\mathcal L_{u,j}$, where $\mathcal L_{u,j}$ denotes candidate $j$'s conditional innovation law, and then forming the treatment-effect contrast.

Validation for stacking follows the weighting and aggregation of $\tau(\mathbf q)$, so its holdouts may differ from those used for point prediction. For a multiyear average, for example, they span a comparable period. We use the notation $D_\ell$ and $V_\ell$ from Appendix~\ref{app:pcpa-point-validation}, but allow different folds for point prediction and predictive calibration.

For each fold, let $\mathbf q_\ell=\{q_{\ell,it}:(i,t)\in V_\ell\}$ be the weights obtained by applying the aggregation rule of $\mathbf q$ to its validation cells. By Assumption~\ref{ass:pcpa-untreated-consistency}, the placebo effect is
\[
\tau_\ell(\mathbf q_\ell)
:=\sum_{(i,t)\in V_\ell}q_{\ell,it}\{Y_{it}-Y_{it}(0)\}=0.
\]
For rate effects, we apply the exposure and unit conversions of Appendix~\ref{app:pcpa-point-validation} to outcomes and predictive draws before forming these aggregates. Let $F_{\ell,\mathbf w}$ denote the predictive distribution function of the corresponding observed-minus-predicted aggregate. We estimate the mixture weights by minimizing the fold-weighted CRPS at zero,
\begin{equation}
\widehat{\mathbf w}\in\arg\min_{\mathbf w\in\Delta_{\mathcal J}}
\sum_{\ell=1}^L a_\ell\operatorname{CRPS}(F_{\ell,\mathbf w},0),
\qquad a_\ell\geq0,\quad\sum_{\ell=1}^L a_\ell=1.
\label{eq:pcpa-stacking-objective}
\end{equation}
\label{eq:crps-stacking-weights}
The continuous ranked probability score is strictly proper for scalar predictive distributions with finite first moments \citep{gneiting2007strictly}. The fold weights $a_\ell$ determine how the validation problems contribute to calibration. The holdout designs and weights for the numerical experiments and policy evaluations are given in Appendices~\ref{app:pcpa-designs} and~\ref{app:pcpa-empirical}.

Once predictions have been generated for each candidate, the weights can be estimated without further predictive simulation. Let $T_{\ell j}^{(b)}$, $b=1,\ldots,B$, be draws of the aggregate contrast for candidate $j$ in fold $\ell$. Define
\begin{equation}
A_{\ell j}=\frac1B\sum_{b=1}^B|T_{\ell j}^{(b)}|,
\qquad
B_{\ell jj'}=\frac1{B^2}\sum_{b=1}^B\sum_{b'=1}^B
 |T_{\ell j}^{(b)}-T_{\ell j'}^{(b')}|.
\label{eq:pcpa-score-coefficients}
\end{equation}
For the empirical mixture,
\begin{equation}
\operatorname{CRPS}(F_{\ell,\mathbf w},0)
=\mathbf A_\ell^{\mathsf T}\mathbf w
 -\tfrac12\mathbf w^{\mathsf T}\mathbf B_\ell\mathbf w,
\label{eq:pcpa-empirical-crps}
\end{equation}
\label{eq:empirical-mixture-crps}
where $\mathbf A_\ell=(A_{\ell j})_{j\in\mathcal J}$ and $\mathbf B_\ell=(B_{\ell jj'})_{j,j'\in\mathcal J}$. These coefficients are calculated once and used throughout the optimization.

For final prediction, a mixture component applies to the complete missing block. We form the treatment-effect contrast within each joint predictive draw and then calculate medians and central predictive intervals from the weighted empirical mixture. This order preserves dependence when aggregating over units or periods.

The placebo-effect CRPS criterion also scores predictions of the untreated aggregate. Let $y_\ell=\sum_{(i,t)\in V_\ell}q_{\ell,it}Y_{it}$ and let $G_{\ell,\mathbf w}$ denote the predictive distribution function of the corresponding untreated aggregate. Translation and reflection invariance of the CRPS give
\[
\operatorname{CRPS}(F_{\ell,\mathbf w},0)
=\operatorname{CRPS}(G_{\ell,\mathbf w},y_\ell).
\]
This identity connects calibration of placebo effects to prediction of the untreated aggregate. Strict propriety makes the true conditional distribution optimal in expected CRPS \citep{gneiting2007strictly}, while selection within the specified mixture family targets its best available approximation in expected score \citep{yao2018stacking}. If the weighted validation criterion converges uniformly over that family to the expected score for predicting the missing untreated aggregate, an approximate minimizer has asymptotically optimal expected score within the family \citep[Section~5.2]{vandervaart1998asymptotic}.

For interval inference, the relevant target is the conditional distribution of the missing untreated aggregate given $\mathcal D_\Omega$. Convergence of the fitted predictive distribution to this law gives asymptotically nominal coverage conditional on $\mathcal D_\Omega$ when the law is continuous and its quantile function is continuous at the interval's endpoint probabilities \citep[Lemma~21.2]{vandervaart1998asymptotic}. Subtracting the predictive endpoints in reverse order from the observed treated aggregate gives an interval for $\tau(\mathbf q)$ with the same coverage event. We assess the resulting coverage in the numerical experiments.

\subsection{Sampling the reference innovations}
\label{app:pcpa-reference-sampling}
\label{app:full-field-residual}

For latent Gaussian models, we obtain $\mathbf V_P(\boldsymbol\theta)$ by conditioning on $\mathbf Z_\Omega(0)$. To use sparse precision calculations, we work with the free Gaussian coefficients after imposing the model's linear constraints. Let $\mathbf x$ collect these coefficients, including those for the additive effects, and write the vectorized latent panel as
\[
\mathbf Z(0)=\mathbf B_{\boldsymbol\theta}\mathbf x,
\qquad
\mathbf x\mid\boldsymbol\theta\sim
\mathcal N(\mathbf0,\mathbf C_{\boldsymbol\theta}),
\qquad
\mathbf Q_{0,\boldsymbol\theta}=\mathbf C_{\boldsymbol\theta}^{-1}.
\]
The matrix $\mathbf B_{\boldsymbol\theta}$ maps these coefficients to the latent predictor, and $\mathbf Q_{0,\boldsymbol\theta}$ is the proper prior precision on the constrained coefficient space. Subscripts $\Omega$ and $\mathcal M$ select rows corresponding to observed and missing cells. Gaussian conditioning gives \citep[Theorem~2.5.1]{anderson2003introduction}
\begin{equation}
\begin{aligned}
\mathbf V_P(\boldsymbol\theta)
={}&\mathbf B_{\mathcal M,\boldsymbol\theta}
\left[\mathbf C_{\boldsymbol\theta}
-\mathbf C_{\boldsymbol\theta}\mathbf B_{\Omega,\boldsymbol\theta}^{\mathsf T}
 (\mathbf B_{\Omega,\boldsymbol\theta}\mathbf C_{\boldsymbol\theta}
   \mathbf B_{\Omega,\boldsymbol\theta}^{\mathsf T})^{-}
 \mathbf B_{\Omega,\boldsymbol\theta}\mathbf C_{\boldsymbol\theta}
\right]
\mathbf B_{\mathcal M,\boldsymbol\theta}^{\mathsf T}.
\end{aligned}
\label{eq:pcpa-reference-sampling-covariance}
\end{equation}
\label{eq:appendix-conditional-covariance}
The generalized inverse accommodates redundant conditioning equations. Conditioning fixes the combined contribution of the coefficients to the observed latent predictor $\mathbf B_{\Omega,\boldsymbol\theta}\mathbf x$, while coefficient perturbations in the null space of $\mathbf B_{\Omega,\boldsymbol\theta}$ remain possible.

At fixed $\boldsymbol\theta$, the conditional covariance is independent of the conditioned value of $\mathbf Z_\Omega(0)$. We can therefore sample centered innovations by conditioning Gaussian coefficient perturbations on a zero observed predictor. For posterior draw $s$, generate
\begin{equation}
\begin{aligned}
\mathbf v^{(s)}&\sim\mathcal N(\mathbf0,\mathbf C_{\boldsymbol\theta^{(s)}})
\quad\text{conditional on}\quad
\mathbf B_{\Omega,\boldsymbol\theta^{(s)}}\mathbf v^{(s)}=\mathbf0,\\
\mathbf e_{J,\mathcal M}^{(s)}
&=\mathbf B_{\mathcal M,\boldsymbol\theta^{(s)}}\mathbf v^{(s)}.
\end{aligned}
\label{eq:pcpa-reference-sampler}
\end{equation}
\label{eq:appendix-joint-innovation-draw}
The resulting innovation has covariance $\mathbf V_P(\boldsymbol\theta^{(s)})$. We impose the conditioning equations through a Schur-complement projection, using sparse solves with $\mathbf Q_{0,\boldsymbol\theta^{(s)}}$ rather than forming the dense coefficient covariance. For the independent reference innovations, we use the same marginal conditional variances and set cross-cell covariances to zero.

Within each posterior draw, the same independent or joint innovation is used at both scale multipliers. Its covariance is evaluated at the model parameters sampled in that draw. Reusing innovations across scales reduces Monte Carlo variation in comparisons of the predictive scores.

\section{Fast cross-validation}
\label{app:pcpa-cavity}
\label{app:cavity-validation}

Cross-validation requires posterior inference after withholding each set of untreated observations. Repeating the full probabilistic-model fit for every fold can be costly. Following leave-group-out methods for latent Gaussian models \citep{liu2025leavegroup,adin2024automatic}, we instead remove the withheld likelihood contributions from a Gaussian approximation to the posterior. This appendix details our approach and provides evidence on accuracy and computational savings compared to refitting the full probabilistic model after deletion.

\subsection{Gaussian representation at the fitted hyperparameters}
\label{app:cavity-fixed-mode}

For validation, we fix $\boldsymbol\theta$ at its posterior mode $\widehat{\boldsymbol\theta}$ estimated from $\mathcal D_\Omega$. Using the coefficient representation of Appendix~\ref{app:pcpa-reference-sampling}, write $\mathbf Q_0=\mathbf Q_{0,\widehat{\boldsymbol\theta}}$ and $\mathbf B=\mathbf B_{\widehat{\boldsymbol\theta}}$. The latent predictor is $\mathbf B\mathbf x$.

For cell $r\in\Omega$, write $\eta_r=\mathbf b_r^\top\mathbf x$, where $\mathbf b_r^\top$ is row $r$ of $\mathbf B$. Represent its Gaussian likelihood contribution, up to a factor independent of $\mathbf x$, by $\exp\{-d_r(\eta_r-\widetilde y_r)^2/2\}$. With $\mathbf D_\Omega^{\mathrm{lik}}= \operatorname{diag}\{d_r:r\in\Omega\}$, the Gaussian posterior approximation has precision $\mathbf Q_\Omega$, information vector $\mathbf h_\Omega$, and mean $\mathbf m_\Omega$ given by
\begin{equation}
\begin{aligned}
\mathbf Q_\Omega
 &=\mathbf Q_0+\mathbf B_\Omega^{\mathsf T}
                  \mathbf D_\Omega^{\mathrm{lik}}\mathbf B_\Omega,\\
\mathbf h_\Omega
 &=\mathbf B_\Omega^{\mathsf T}
                  \mathbf D_\Omega^{\mathrm{lik}}\widetilde{\mathbf y}_\Omega,
\qquad
\mathbf m_\Omega=\mathbf Q_\Omega^{-1}\mathbf h_\Omega.
\end{aligned}
\label{eq:pcpa-cavity-full-system}
\end{equation}
\label{eq:cavity-full-system}
For Gaussian observations, $d_r$ is the observation precision and $\widetilde y_r=Y_r$, giving the exact posterior at the fixed hyperparameters. For non-Gaussian observations, we approximate each log-likelihood contribution $\ell_r$ by a quadratic expansion at the predictor $\widehat\eta_r$ obtained using all untreated observations. The corresponding working precision and response are \citep{rue2009approximate,liu2025leavegroup}
\begin{equation}
d_r=-\ell_r''(\widehat\eta_r),
\qquad
\widetilde y_r=\widehat\eta_r+
                 \frac{\ell_r'(\widehat\eta_r)}{d_r},
\qquad d_r>0.
\label{eq:pcpa-cavity-likelihood}
\end{equation}
\label{eq:cavity-working-general}
In the Poisson analyses, $\widehat\mu_r=E_r\exp(\widehat\eta_r)$, $d_r=\widehat\mu_r$, and $\widetilde y_r=\widehat\eta_r+(Y_r-\widehat\mu_r)/\widehat\mu_r$.

\subsection{Withholding observations}
\label{app:cavity-deletion}
\label{app:cavity-approximation}

By Assumption~\ref{ass:pcpa-conditional-observations}, the log likelihood is additive over cells. Removing the contributions from $D_\ell$ in Equation~\eqref{eq:pcpa-cavity-full-system} gives
\begin{equation}
\begin{aligned}
\mathbf Q_{-\ell}
 &=\mathbf Q_\Omega-\mathbf B_{D_\ell}^{\mathsf T}
                        \mathbf D_{D_\ell}^{\mathrm{lik}}\mathbf B_{D_\ell},\\
\mathbf h_{-\ell}
 &=\mathbf h_\Omega-\mathbf B_{D_\ell}^{\mathsf T}
                        \mathbf D_{D_\ell}^{\mathrm{lik}}
                        \widetilde{\mathbf y}_{D_\ell}.
\end{aligned}
\label{eq:pcpa-cavity-deletion}
\end{equation}
\label{eq:cavity-downdate}
For Gaussian observation models, subtracting the withheld likelihood contributions gives the exact posterior conditional on the fixed hyperparameters. For non-Gaussian models, we use the same subtraction with the quadratic likelihood expansion from the original fit, without recomputing it after deletion. In both cases, we keep the hyperparameters fixed at their full-data estimates, rather than updating their posterior distribution after withholding observations.

To obtain joint draws from this approximate posterior, factor $\mathbf Q_{-\ell}=\mathbf L_{-\ell}\mathbf L_{-\ell}^{\mathsf T}$ and generate
\begin{equation}
\mathbf x_{-\ell}^{(b)}
=\mathbf Q_{-\ell}^{-1}\mathbf h_{-\ell}
 +\mathbf L_{-\ell}^{-\mathsf T}\boldsymbol\varepsilon^{(b)},
\qquad
\boldsymbol\varepsilon^{(b)}\sim\mathcal N(\mathbf0,\mathbf I).
\label{eq:pcpa-cavity-draws}
\end{equation}
\label{eq:cavity-draw}
The input to $c_k$ is $\mathbf B_{\Omega\setminus D_\ell}\mathbf x_{-\ell}^{(b)}$. We evaluate the point and predictive validation criteria in Appendix~\ref{app:pcpa-predictive} using the resulting imputations.

\subsection{Comparison with refitting the probabilistic model}
\label{app:cavity-refit-check}

We assess the approximation by refitting the probabilistic model after withholding observations. Refitting at fixed hyperparameters isolates the error from the Gaussian likelihood approximation; re-estimating the hyperparameters also assesses the effect of fixing their original values. We make both comparisons for nine empirical holdouts: the first fold used to compare point predictions for each application and probabilistic model, two opioid-mortality holdouts assigning the adoption schedule to untreated states, and two nineteen-year Vision Zero holdouts, including Croatia in 1990--2008.

At fixed hyperparameters, the Gaussian calculations agree to numerical precision (Table~\ref{tab:pcpa-cavity-benchmarks}, Panel A). The Poisson comparisons also give small differences in posterior means and marginal standard deviations on the log-rate scale. Re-estimating hyperparameters changes some standard deviations more noticeably. Using predictions from direct refitting for the Croatian validation holdout also has little effect on the final predictive interval for Sweden.

We also measure the time needed to compute each posterior and obtain 512 joint draws of the latent panel, excluding causal estimation and predictive stacking. On an Intel Core Ultra 9 285K processor, with serial execution and one INLA and BLAS thread, the one-time setup for the cavity approximation takes 0.0099--0.0270 seconds per model. Subsequent folds take 0.0235--0.244 seconds. Refitting takes 1.86--13.00 seconds at fixed hyperparameters and 12.4--52.5 seconds when hyperparameters are re-estimated. Including setup across the tested holdouts, the cavity approximation reduces posterior computation time by factors of 110--800 relative to refitting with hyperparameters re-estimated. Panel B reports the timings for each holdout.

\begin{table}[!tbp]
\centering
\begin{threeparttable}
\caption{Accuracy and computing time of the cavity approximation.}
\label{tab:pcpa-cavity-benchmarks}
\label{tab:cavity-refit-check}
\small\setstretch{1}
\setlength{\tabcolsep}{0pt}
\begin{tabular*}{\linewidth}{@{}>{\raggedright\arraybackslash}p{96pt}
@{\hspace{6pt}}>{\raggedright\arraybackslash}p{\dimexpr\linewidth-330pt\relax}
@{\hspace{10pt}}S[table-format=1.1e-1,table-column-width=50pt]
@{\hspace{4pt}}S[table-format=1.1e-1,table-column-width=50pt]
@{\hspace{10pt}}S[table-format=1.1e-1,table-column-width=50pt]
@{\hspace{4pt}}S[table-format=1.1e-1,table-column-width=50pt]@{}}
\toprule
\multicolumn{6}{l}{\textit{Panel A. Differences in posterior means and standard deviations}} \\
\addlinespace[3pt]
& & \multicolumn{2}{c}{Fixed hyperparameters} & \multicolumn{2}{c}{Re-estimated} \\
\cmidrule(lr){3-4}\cmidrule(lr){5-6}
Application / model & Holdout & {Means} & {SDs} & {Means} & {SDs} \\
\midrule
California / Nonsp. & Alabama, 1989--2000 & {\(<10^{-8}\)} & {\(<10^{-8}\)} & 7.2e-03 & 6.9e-04 \\
California / Spatial & Alabama, 1989--2000 & {\(<10^{-8}\)} & {\(<10^{-8}\)} & 1.6e-03 & 1.8e-03 \\
Cannabis / Nonsp. & Point prediction & 5.0e-03 & 6.2e-05 & 4.4e-03 & 1.1e-03 \\
Cannabis / Nonsp. & Reassigned adoption schedule & 2.2e-03 & 2.1e-04 & 1.9e-03 & 8.2e-04 \\
Cannabis / Spatial & Point prediction & 9.6e-04 & 6.6e-05 & 1.5e-03 & 2.4e-03 \\
Cannabis / Spatial & Reassigned adoption schedule & 6.8e-04 & 1.7e-04 & 9.6e-04 & 7.1e-04 \\
Vision Zero / Nonsp. & Point prediction & 5.0e-04 & 9.2e-06 & 5.2e-04 & 1.3e-03 \\
Vision Zero / Nonsp. & 1997 holdout & 4.1e-04 & 1.1e-06 & 1.2e-03 & 4.7e-03 \\
Vision Zero / Nonsp. & Croatia, 1990--2008 & 4.7e-04 & 1.6e-05 & 1.1e-03 & 4.1e-03 \\
\bottomrule
\end{tabular*}
\par\medskip
\begin{tabular*}{\linewidth}{@{}>{\raggedright\arraybackslash}p{96pt}
@{\hspace{6pt}}>{\raggedright\arraybackslash}p{\dimexpr\linewidth-338pt\relax}
@{\hspace{10pt}}S[table-format=1.3,table-column-width=48pt]
@{\hspace{4pt}}S[table-format=2.2,table-column-width=32pt]
@{\hspace{10pt}}S[table-format=2.2,table-column-width=64pt]
@{\hspace{4pt}}S[table-format=4.0,table-column-width=64pt]@{}}
\multicolumn{6}{l}{\textit{Panel B. Time to obtain 512 joint draws of the latent panel}} \\
\addlinespace[3pt]
& & & \multicolumn{2}{c}{Direct refit (s)} & \\
\cmidrule(lr){4-5}
Application / model & Holdout & {Cavity (s)} & {Fixed} & {Re-estimated} & {Speedup} \\
\midrule
California / Nonsp. & Alabama, 1989--2000 & 0.099 & 2.54 & 12.36 & 120 \\
California / Spatial & Alabama, 1989--2000 & 0.244 & 12.95 & 52.50 & 220 \\
Cannabis / Nonsp. & Point prediction & 0.091 & 2.45 & 33.49 & 370 \\
Cannabis / Nonsp. & Reassigned adoption schedule & 0.080 & 2.44 & 32.84 & 410 \\
Cannabis / Spatial & Point prediction & 0.228 & 12.82 & 44.98 & 200 \\
Cannabis / Spatial & Reassigned adoption schedule & 0.222 & 12.79 & 45.13 & 200 \\
Vision Zero / Nonsp. & Point prediction & 0.025 & 1.86 & 22.53 & 900 \\
Vision Zero / Nonsp. & 1997 holdout & 0.023 & 1.86 & 22.34 & 950 \\
Vision Zero / Nonsp. & Croatia, 1990--2008 & 0.025 & 1.86 & 22.38 & 880 \\
\bottomrule
\end{tabular*}
\begin{tablenotes}[flushleft]
\footnotesize
\item Notes: Panel A reports root mean squared differences between cavity and directly refitted posterior means and marginal standard deviations over withheld cells. Comparisons use standardized sales for tobacco control and log rates for the two count-data applications. Direct refits either fix hyperparameters at their full-data values or re-estimate them. Nonsp.\ and Spatial denote nonspatial and spatial probabilistic models. Panel B reports median computation times over three runs for posterior construction and 512 joint latent-panel draws. Times exclude the initial model fit and the one-time cavity setup. Speedup is the re-estimated-hyperparameter refit time divided by the cavity time; the overall speedup reported in the text includes cavity setup. Causal estimation and predictive stacking are excluded throughout.
\end{tablenotes}
\end{threeparttable}
\end{table}

\section{Causal panel estimators: implementation details}
\label{app:pcpa-estimators}
\label{app:fast-sc}

The examples below show how several established causal panel estimators can be applied within the construction in Section~\ref{subsec:pcpa-draw-wise}. We give the fitting rules used in the numerical experiments and empirical illustrations, including the numerical details needed for repeated estimation across posterior draws. Study-specific predictors and tuning choices are given with the analyses in Appendices~\ref{app:pcpa-designs} and~\ref{app:pcpa-empirical}.

\subsection{Interactive fixed effects}
\label{ex:interactive-fixed-effects}

An interactive fixed-effects model represents the untreated panel by
\[
Z_{it}(0)\approx\alpha_i+\delta_t+
\boldsymbol\lambda_i^{\mathsf T}\mathbf f_t,
\qquad \boldsymbol\lambda_i,\mathbf f_t\in\mathbb R^R,
\]
where $R$ is the factor rank. For each latent input panel $\mathbf z_\Omega$, we use generalized synthetic controls \citep{xu2017generalized} to estimate the period effects and common factor paths from never-treated controls, and each treated unit's intercept and loadings from its pretreatment history. The imputation is
\[
c_{k,it}(\mathbf z_\Omega)
=
\hat\alpha_i+\hat\delta_t+
\hat{\boldsymbol\lambda}_i^\top\hat{\mathbf f}_t,
\qquad (i,t)\in\mathcal M,
\]
with all fitted quantities recomputed for each input panel. Observed time-varying covariates can also enter the model; their estimation in the opioid-mortality analysis is described in Appendix~\ref{subsec:pcpa-cannabis}.\footnote{A related interactive fixed-effects estimator uses all untreated cells to estimate the common factor structure, including treated units' pretreatment observations; see \citet{gobillon2016regional}.}

\subsection{Matrix completion}
\label{ex:matrix-completion}

For matrix completion, we estimate additive unit and period effects with a nuclear-norm penalized low-rank component \citep{athey2021matrix}. Given a latent input panel $\mathbf z_\Omega$, we solve
\[
(\widehat{\mathbf L},\widehat{\boldsymbol\gamma},
 \widehat{\boldsymbol\delta})
\in\arg\min_{\mathbf L,\boldsymbol\gamma,\boldsymbol\delta}
\left\{
\frac1{|\Omega|}\sum_{(i,t)\in\Omega}
(z_{it}-L_{it}-\gamma_i-\delta_t)^2
+\lambda\|\mathbf L\|_*\right\}.
\]
The imputation is $c_{k,it}(\mathbf z_\Omega)=\widehat L_{it} +\widehat\gamma_i+\widehat\delta_t$ on the missing cells. The opioid-mortality analysis also includes policy covariates and uses the penalty scaling described in Appendix~\ref{subsec:pcpa-cannabis}.

\subsection{Computing synthetic-control weights across posterior draws}
\label{app:pcpa-scm-computation}

Synthetic-control fits across posterior draws share the control pool and pretreatment periods. We use the solution at the posterior mean panel to initialize a control set for the draw-specific optimizations.

For each treated unit and posterior draw, we assemble the matching predictors, including any fixed covariates and predictor weights. We divide the treated and control inputs by their common maximum absolute magnitude and denote the resulting vector and matrix by $\widetilde{\mathbf y}$ and $\mathbf X$. We solve
\begin{equation}
\begin{aligned}
\widehat{\boldsymbol\omega}
&\in\arg\min_{\boldsymbol\omega\in\Delta_{\mathcal C}}
\left\{\tfrac12\|\widetilde{\mathbf y}-\mathbf X\boldsymbol\omega\|_2^2
      +\tfrac{\lambda_{\mathrm{sc}}}{2}\|\boldsymbol\omega\|_2^2\right\},\\
\Delta_{\mathcal C}
&=\{\boldsymbol\omega\geq0:\mathbf1^{\mathsf T}\boldsymbol\omega=1\},
\qquad\lambda_{\mathrm{sc}}=10^{-6}.
\end{aligned}
\label{eq:pcpa-scm-ridge-objective}
\end{equation}
\label{eq:sc-drawwise-qp}
The small fixed ridge gives a unique minimizer. It is not selected by validation and is distinct from the augmentation penalty below.

We first solve this problem over the full control pool using the posterior mean panel. Controls whose weights exceed the numerical active-weight threshold form the initial set for each posterior draw. After solving on this set, we check simplex feasibility and the Karush--Kuhn--Tucker conditions over all controls \citep[Sec.~5.5.3]{boyd2004convex}. When the check fails, we solve and refine the full problem until the same conditions hold to numerical tolerance. The initial control set affects computation only; it does not restrict the accepted draw-specific solution. We refine the final weights to a scaled KKT residual no greater than $10^{-12}$.

\subsection{Ridge-augmented synthetic controls}
\label{app:pcpa-ascm-implementation}

Ridge-augmented synthetic controls use a regression adjustment for imperfect pretreatment matching \citep{ben2021augmented}. For a latent input panel $\mathbf z_\Omega$, let $\hat\omega_{ij}$ be unit $i$'s synthetic-control weights and let $\hat m_{it}$ be its ridge-regression prediction from pretreatment outcomes, fitted using the controls. The augmented imputation is
\[
\begin{aligned}
c_{k,it}(\mathbf z_\Omega)
&=
\sum_{j\in\mathcal C}\hat\omega_{ij}z_{jt}
+\hat m_{it}
-\sum_{j\in\mathcal C}\hat\omega_{ij}\hat m_{jt},
\qquad (i,t)\in\mathcal M.
\end{aligned}
\]
Both the control weights and regression predictions are re-estimated for each input panel. With ridge regression, this imputation can also be written as a weighted combination of control outcomes, allowing negative weights and extrapolation beyond their convex hull.

For each input panel, we use pretreatment outcomes as predictors, standardizing each period by its mean and standard deviation across control units. The same transformation is applied to the treated unit's predictors. We set the ridge penalty to $\lambda=r s_G$, where $s_G$ is the mean of the positive eigenvalues of the control Gram matrix, formed from inner products of the standardized control predictor vectors. This scaling expresses the penalty relative to the magnitude of the predictor matrix, with $r$ controlling the degree of regularization. The augmented weights are applied to the original, unstandardized control trajectories. We fix $r=1$ in the Poisson experiment and select it by cross-validation in the Vision Zero analysis.

\section{Statistical properties and proofs}
\label{app:pcpa-theory}
\label{app:proofs-supporting}

Section~\ref{sec:pcpa-theory} separates the accuracy of the latent input from the recovery and stability properties of the causal estimator. We first prove the posterior-consistency result and explain how its conditions relate to synthetic controls, interactive fixed effects, and matrix completion. We then prove the pooling-risk identity and the Bayes-risk bound, and examine how Gaussian pooling changes the factor space estimated from the control panel.

Posterior consistency here concerns $\mathbf m_0(\mathbf z_\Omega^\star)$. Predictive interval coverage requires additional distributional accuracy, as discussed in Appendix~\ref{app:pcpa-stacking}.

\subsection{Posterior consistency and contraction rates}
\label{app:pcpa-transfer-details}
\label{app:proof-imputation-transfer}

\begin{proof}[Proof of Theorem~\ref{thm:pcpa-transfer}]
For any $\mathbf z$ with $d_\Omega(\mathbf z,\mathbf z_\Omega^\star)\leq r$, the triangle inequality gives
\[
\begin{aligned}
d_{\mathcal M}\{c_k(\mathbf z),\mathbf m_0(\mathbf z_\Omega^\star)\}
&\leq
 d_{\mathcal M}\{c_k(\mathbf z),c_k(\mathbf z_\Omega^\star)\}+b_k\\
&\leq b_k+\zeta_k(r).
\end{aligned}
\]
The event on the left of Equation~\eqref{eq:pcpa-transfer-bound} is therefore contained in the event on the right. Taking posterior probabilities conditional on $\mathcal D_\Omega$ proves the bound.

Set $a_n=b_{k,n}+\zeta_{k,n}(r_n)$. By Assumption~\ref{ass:pcpa-imputation-consistency}(ii)--(iii), $a_n=o_{P_0}(1)$. For any fixed $\epsilon>0$, with $P_0$-probability approaching one, $a_n\leq\epsilon$. Applying the bound at $r=r_n$ and using part (i) then proves posterior consistency.

Under the theorem's local Lipschitz condition, $a_n\leq b_{k,n}+L_{k,n}r_n=O_{P_0}(r_n)$. Thus, for any deterministic $M_n\to\infty$, $P_0(a_n\leq M_nr_n)\to1$. Applying the same posterior bound at $r_n$ gives contraction at rate $r_n$.
\end{proof}

\subsection{Recovery and stability for specific estimators}
\label{app:pcpa-recovery}

We consider the requirements $b_{k,n}=o_{P_0}(1)$ and $\zeta_{k,n}(r_n)=o_{P_0}(1)$ in Assumption~\ref{ass:pcpa-imputation-consistency}\allowbreak(ii)--(iii) for the estimators in Appendix~\ref{app:pcpa-estimators}. The arguments below begin with a specified estimator on a fixed panel. Their bounds must vanish in the normalized distances along the chosen asymptotic sequence to establish consistency. The contraction-rate conclusion additionally requires $b_{k,n}=O_{P_0}(r_n)$ and $L_{k,n}=O_{P_0}(1)$.

\subsubsection{Synthetic controls}
\label{app:pcpa-recovery-synthetic-controls}
\label{ex:synthetic-control-stability}

Suppose that $\mathbf z_\Omega^\star$ and $\mathbf m_0(\mathbf z_\Omega^\star)$ share the additive factor representation $\alpha_i+\delta_t+ \boldsymbol\lambda_i^{\mathsf T}\mathbf f_t$. With weights summing to one, the treated-minus-control difference is
\[
\left(\alpha_i-\sum_{j\in\mathcal C}\omega_j\alpha_j\right)
+
\left(\boldsymbol\lambda_i-
      \sum_{j\in\mathcal C}\omega_j\boldsymbol\lambda_j\right)^{\mathsf T}
\mathbf f_t.
\]
If the treated coefficient vector lies in the control convex hull and the pretreatment factor design, including an intercept, has full column rank, exact pretreatment matching at the true latent input identifies this balance. It therefore gives $b_k=0$ \citep{abadie2010synthetic,abadie2021using}.

Local stability also depends on how the fitted weights change. Write $\boldsymbol\omega_i(\mathbf z)$ for the selected weight vector and $\mathbf z_{\mathcal C,t}$ for the vector of control outcomes at period $t$ extracted from the latent input $\mathbf z$. Superscript $\star$ denotes the corresponding quantities at the true input. For each missing cell,
\[
\begin{aligned}
\bigl|c_{k,it}(\mathbf z)-c_{k,it}(\mathbf z_\Omega^\star)\bigr|
&\leq
\|\mathbf z_{\mathcal C,t}-\mathbf z_{\mathcal C,t}^\star\|_2\\
&\quad+
\|\mathbf z_{\mathcal C,t}^\star\|_2
\|\boldsymbol\omega_i(\mathbf z)-
  \boldsymbol\omega_i(\mathbf z_\Omega^\star)\|_2,
\end{aligned}
\]
where the first term uses $\|\boldsymbol\omega_i(\mathbf z)\|_2\leq1$ on the weight simplex. The bound separates the direct change in control outcomes from the change in their fitted weights. A unique solution satisfying local quadratic growth permits control of the latter through optimization perturbation results \citep[Sec.~4.4.1]{bonnans2000perturbation}. Aggregating these inequalities in $d_{\mathcal M}$ bounds $\zeta_k(r)$. Assumption~\ref{ass:pcpa-imputation-consistency}(iii) follows when the resulting bound at $r=r_n$ is $o_{P_0}(1)$.

For the synthetic-control estimator with the fixed ridge penalty in Appendix~\ref{app:pcpa-scm-computation}, $b_k$ measures imputation error at the penalized solution. The ridge stabilizes the optimization but may leave nonzero oracle error, as can imperfect matching.

\subsubsection{Interactive fixed effects}
\label{app:pcpa-recovery-ife}
\label{ex:interactive-fixed-effects-stability}

For generalized synthetic controls, the controls must identify the common factor space and each treated unit's pretreatment history must identify its loadings. Treatment may depend on the factor structure, but must be exogenous to the remaining idiosyncratic shocks \citep{xu2017generalized}. Under strong factors and the requisite moment and dependence conditions, recovery of the factor component improves as the number of controls and pretreatment periods increases \citep{bai2009panel,xu2017generalized}. If that component equals $\mathbf m_0(\mathbf z_\Omega^\star)$, consistent recovery in $d_{\mathcal M}$ gives $b_{k,n}=o_{P_0}(1)$.

A positive gap between the $R$th and $(R+1)$th singular values of the control residual panel controls perturbations of the estimated factor space \citep{stewart1990matrix}. Full column rank of the treated unit's pretreatment regression design, including its intercept, then controls perturbations of the fitted loadings. Together these conditions provide a local bound on $\zeta_k(r)$; part (iii) of Assumption~\ref{ass:pcpa-imputation-consistency} requires $\zeta_{k,n}(r_n)=o_{P_0}(1)$. We relate factor-space perturbations to Gaussian pooling in Appendix~\ref{app:pcpa-low-rank-details}.

\subsubsection{Matrix completion}
\label{app:pcpa-recovery-matrix-completion}
\label{ex:matrix-completion-stability}

For matrix completion, low rank must be accompanied by enough untreated observations to recover the missing conditional means. In the causal model of \citet{athey2021matrix}, adoption can depend on the systematic component but is conditionally independent of the idiosyncratic shocks. When their recovery bounds apply to the specified estimator and the target $\mathbf m_0(\mathbf z_\Omega^\star)$, vanishing error in $d_{\mathcal M}$ gives $b_{k,n}=o_{P_0}(1)$.\footnote{The bounds in \citet{athey2021matrix} concern a bounded-entry low-rank model without additional fixed effects and measure error over the full panel. Applying them here requires the corresponding control over the missing cells.} A fixed positive nuclear-norm penalty can leave nonzero oracle error through shrinkage even when the low-rank representation is correct.

For local stability, fix the observation pattern and a full-rank design for the additive effects and any covariates. Profiling out these coefficients leaves nuclear-norm regularized least squares with a fixed linear operator. For a positive penalty, \citet[Theorem~4.7]{nghia2025geometric} establishes local Lipschitz continuity when the optimizer is unique throughout a neighborhood of the input and penalty. Reconstructing the unpenalized coefficients is linear in the input and fitted matrix, so the counterfactual imputation inherits this stability. With the penalty ratio in Appendix~\ref{subsec:pcpa-cannabis} fixed, the penalty is itself a locally Lipschitz function of the input, since residualization is linear and the largest singular value is Lipschitz. The same argument therefore covers this scaling rule wherever the reference singular value is positive and the neighborhood uniqueness condition holds.

\subsection{Pooling and imputation risk}
\label{app:pcpa-pooling-proof}
\label{app:proof-pooling-risk}

Proposition~\ref{prop:pcpa-pooling} compares risk conditional on $\mathbf Z_\Omega(0)=\mathbf z_\Omega^\star$, whereas Proposition~\ref{prop:pcpa-bayes-risk} averages over the joint distribution of $\mathbf Z_\Omega(0)$ and $\mathcal D_\Omega$ under $P_0$.

\begin{proof}[Proof of Proposition~\ref{prop:pcpa-pooling}]
All expectations and variances are conditional on $\mathbf Z_\Omega(0)=\mathbf z_\Omega^\star$. Using $b_k=0$ and $\widehat{\mathbf z}_P=\mathbf S\widehat{\mathbf z}_U$, the unpooled and pooled imputation errors are
\[
\begin{aligned}
c_k(\widehat{\mathbf z}_U)-\mathbf m_0(\mathbf z_\Omega^\star)
  &=\mathbf J_k\boldsymbol\varepsilon,\\
c_k(\widehat{\mathbf z}_P)-\mathbf m_0(\mathbf z_\Omega^\star)
  &=\mathbf J_k(\mathbf S-\mathbf I)\mathbf z_\Omega^\star
    +\mathbf J_k\mathbf S\boldsymbol\varepsilon.
\end{aligned}
\]
The smoothing term in the second line is fixed, whereas the noise term has conditional mean zero. Its cross product with the smoothing term therefore has expectation zero. Taking squared norms and expectations gives
\[
\begin{aligned}
|\mathcal M|\mathcal R_k(\widehat{\mathbf z}_U\mid\mathbf z_\Omega^\star)
 &=\operatorname{tr}(\mathbf J_k\boldsymbol\Sigma\mathbf J_k^{\mathsf T}),\\
|\mathcal M|\mathcal R_k(\widehat{\mathbf z}_P\mid\mathbf z_\Omega^\star)
 &=\|\mathbf J_k(\mathbf S-\mathbf I)\mathbf z_\Omega^\star\|_2^2\\
 &\quad+\operatorname{tr}(\mathbf J_k\mathbf S\boldsymbol\Sigma
                         \mathbf S^{\mathsf T}\mathbf J_k^{\mathsf T}).
\end{aligned}
\]
Subtracting and using $\mathbf H_k=\mathbf J_k^{\mathsf T}\mathbf J_k$ and cyclicity of the trace gives Equation~\eqref{eq:pcpa-pooling-identity}.
\end{proof}

\begin{proof}[Proof of Proposition~\ref{prop:pcpa-bayes-risk}]
Write $\mathbf Z=\mathbf Z_\Omega(0)$ and let
\[
\mathbf a_0(\mathcal D_\Omega)
=
\mathbb E_0\{\mathbf m_0(\mathbf Z)\mid\mathcal D_\Omega\}.
\]
For any predictor $\mathbf a$ with finite risk, expanding the squared error and conditioning on $\mathcal D_\Omega$ gives
\[
\mathcal B(\mathbf a)
=
\mathcal B(\mathbf a_0)
+
\mathbb E_0\!\left[
 d_{\mathcal M}^2\{
  \mathbf a(\mathcal D_\Omega),
  \mathbf a_0(\mathcal D_\Omega)
 \}
\right],
\]
since $\mathbb E_0\{ \mathbf m_0(\mathbf Z)-\mathbf a_0(\mathcal D_\Omega) \mid\mathcal D_\Omega\}=\mathbf0$. Hence $\mathcal B(\mathbf a_0)=\inf_{\mathbf a}\mathcal B(\mathbf a)$.

Equality of the conditional laws makes the posterior expectation defining $\bar{\mathbf c}_k$ equal to $\mathbb E_0\{c_k(\mathbf Z)\mid\mathcal D_\Omega\}$. Therefore,
\[
\bar{\mathbf c}_k(\mathcal D_\Omega)
-\mathbf a_0(\mathcal D_\Omega)
=
\mathbb E_0\{
 c_k(\mathbf Z)-\mathbf m_0(\mathbf Z)
 \mid\mathcal D_\Omega
\}.
\]
Applying the squared-error identity and Jensen's inequality yields
\[
\begin{aligned}
\mathcal B(\bar{\mathbf c}_k)
-\inf_{\mathbf a}\mathcal B(\mathbf a)
&=
\frac{1}{|\mathcal M|}
\mathbb E_0
\left\|
 \mathbb E_0\{
  c_k(\mathbf Z)-\mathbf m_0(\mathbf Z)
  \mid\mathcal D_\Omega
 \}
\right\|_2^2\\
&\leq
\mathbb E_0\!\left[
 d_{\mathcal M}^2\{
  c_k(\mathbf Z),\mathbf m_0(\mathbf Z)
 \}
\right].
\end{aligned}
\]
\end{proof}

\subsection{Gaussian pooling and recovery of the factor space}
\label{app:pcpa-low-rank-details}
\label{app:pooling-low-rank}

Pooling can also change the factor space used for counterfactual imputation. We examine this effect in the fully observed control block after removing additive unit and period components.

Let $\mathbf Z_{\mathcal C}^\star$ be the true control block, of rank $R$, and let $\mathbf F$ contain an orthonormal basis for its temporal factor space. For a posterior draw, write $\boldsymbol\Delta_{\mathcal C}^{(s)} =\mathbf Z_{\mathcal C}^{(s)}-\mathbf Z_{\mathcal C}^\star$ and $\epsilon^{(s)}=\|\boldsymbol\Delta_{\mathcal C}^{(s)}\|_{\mathrm{op}}$. The operator norm measures the perturbation relevant to the singular space, rather than average cellwise error. Let $\widehat{\mathbf F}^{(s)}$ contain the leading $R$ right singular vectors of the sampled block. Because the factor basis can be rotated without changing the fitted space, we compare the projection matrices. When $\epsilon^{(s)}<\sigma_R(\mathbf Z_{\mathcal C}^\star)/2$, singular-space perturbation bounds give
\begin{equation}
\left\|\widehat{\mathbf F}^{(s)}\widehat{\mathbf F}^{(s)\mathsf T}
      -\mathbf F\mathbf F^{\mathsf T}\right\|_{\mathrm{op}}
\leq
\frac{\epsilon^{(s)}}
     {\sigma_R(\mathbf Z_{\mathcal C}^\star)-\epsilon^{(s)}}.
\label{eq:pcpa-factor-distance}
\end{equation}
\label{eq:factor-recovery-bound}
Here $\sigma_R$ is the smallest nonzero singular value of the true block \citep{stewart1990matrix}. The bound shows why perturbations must be small relative to the strength of the weakest retained factor.

For the smoothing calculation, consider a zero-mean Gaussian working model with fixed covariance $\mathbf K_{ST}=\mathbf K_T\otimes\mathbf K_S$ and independent Gaussian observation errors of variance $\sigma^2$. Suppressing the control-block subscript, write the observed matrix as $\mathbf Z^\star+\mathbf E$ and let $\overline{\mathbf Z}$ denote the posterior mean of the latent matrix. In vectorized form,
\[
\operatorname{vec}(\overline{\mathbf Z})
=
\mathbf S\operatorname{vec}(\mathbf Z^\star+\mathbf E),
\qquad
\mathbf S=\mathbf K_{ST}
(\mathbf K_{ST}+\sigma^2\mathbf I)^{-1}.
\]

Let $\mathbf U_S$ and $\mathbf U_T$ contain orthonormal eigenvectors of $\mathbf K_S$ and $\mathbf K_T$, with eigenvalues $\kappa_{S,a}$ and $\kappa_{T,b}$. Define
\[
\begin{aligned}
\widetilde{\mathbf Z}^{\star}
&=\mathbf U_S^\top\mathbf Z^\star\mathbf U_T,
&\qquad
\widetilde{\mathbf E}
&=\mathbf U_S^\top\mathbf E\mathbf U_T,\\
\widetilde{\overline{\mathbf Z}}
&=\mathbf U_S^\top\overline{\mathbf Z}\mathbf U_T.
\end{aligned}
\]
In these coordinates, the posterior mean satisfies
\begin{equation}
\widetilde{\overline Z}_{ab}
=h_{ab}(\widetilde Z^\star_{ab}+\widetilde E_{ab}),
\qquad
h_{ab}=\frac{\kappa_{S,a}\kappa_{T,b}}
             {\kappa_{S,a}\kappa_{T,b}+\sigma^2}.
\label{eq:pcpa-spectral-multiplier}
\end{equation}
\label{eq:spacetime-spectral-filter}
The corresponding error is $\widetilde{\overline Z}_{ab}-\widetilde Z^\star_{ab} =-(1-h_{ab})\widetilde Z^\star_{ab}+h_{ab}\widetilde E_{ab}$. Pooling removes some signal along with observation noise. Factor recovery can improve when the signal lies mainly in coordinates with $h_{ab}$ close to one and noise is suppressed elsewhere. Shrinking coordinates that carry important factor variation can instead impair recovery. The multipliers $h_{ab}$ need not factor into separate spatial and temporal multipliers, so Gaussian pooling need not preserve matrix rank.

For posterior draws at the same fixed parameters, the error also includes variation around $\overline{\mathbf Z}_{\mathcal C}$,
\[
\boldsymbol\Delta_{\mathcal C}^{(s)}
=
(\overline{\mathbf Z}_{\mathcal C}-\mathbf Z_{\mathcal C}^\star)
+
(\mathbf Z_{\mathcal C}^{(s)}-\overline{\mathbf Z}_{\mathcal C}).
\]
Both terms contribute to $\epsilon^{(s)}$ in Equation~\eqref{eq:pcpa-factor-distance}.

\section{Numerical experiments: designs and results}
\label{app:pcpa-designs}
\label{app:numerical-details}

This appendix gives the designs and full results of the three numerical experiments summarized in Section~\ref{subsec:pcpa-numerical-evidence}.

\subsection{Counterfactual imputation accuracy}
\label{app:experiment1-design}
\label{subsec:pcpa-experiment-accuracy}
\label{subsec:experiment-center}

Reducing observation noise can make the common factor structure easier to estimate, but spatial pooling may also suppress variation needed for counterfactual imputation. We examine this trade-off by varying both observation noise and the spatial structure of the factor loadings. We apply generalized synthetic controls (GSC; \citealp{xu2017generalized}) to each observed or sampled latent panel.

\subsubsection{Design and implementation}

We generate panels on a $6\times10$ rook lattice with $N=60$ units, $T=30$ periods, and $T_0=20$ pretreatment periods. Write $Z_{it}$ for the latent untreated outcome in this experiment. The panel follows
\[
Z_{it}
=
\alpha_i+\delta_t+
\frac{0.90}{\sqrt{2}}
\sum_{r=1}^{2}\lambda_{ir}(\gamma)f_{rt}.
\]
The unit effects are Gaussian with covariance $0.35^2\mat K_I$, where $\mat K_I=\{N/(N-1)\}(\mat I-\vect1\vect1\trans/N)$. This covariance imposes a zero sum on otherwise independent unit variation and has unit average marginal variance. The period effect follows a stationary AR(1) process with correlation $0.6$ and marginal standard deviation $0.25$. The two factor paths are independent stationary AR(1) processes with correlation $0.7$ and unit marginal variance. We discard the first 100 periods of each generated path.

We vary spatial structure in the loadings by combining a smooth spatial component with unstructured variation. Both components are scaled to unit average marginal variance. For the smooth component, let $d_j$ and $\vect v_j$, $j=1,\ldots,6$, be the six smallest positive eigenvalues of the lattice Laplacian and their orthonormal eigenvectors. We use
\[
\mat K_{\mathrm{smooth}}
=
\frac{N}{\sum_{j=1}^{6}d_j^{-1}}
\sum_{j=1}^{6}d_j^{-1}\vect v_j\vect v_j\trans.
\]
The six low-frequency modes produce smooth loadings. The fitted spatial model instead uses the full lattice graph, so its covariance differs from that used to generate the spatial component of the loadings. For each factor, we independently draw $\vect S_r\sim\N(\vect0,\mat K_{\mathrm{smooth}})$ and $\vect U_r\sim\N(\vect0,\mat K_I)$, then set
\[
\vect\lambda_r(\gamma)
=
\sqrt{\gamma}\vect S_r+\sqrt{1-\gamma}\vect U_r,
\qquad
\gamma\in\{0,0.5,1\}.
\]
These values give unstructured, mixed, and fully spatially structured loadings, respectively.

Treatment assignment depends on the factor loadings, but is held fixed across settings within each replication. We choose the ten treated units using reference loadings $\lambda_{ir}^{\mathrm{ref}}$ with equal structured and unstructured shares. Their structured component uses the full scaled Besag covariance. The units with the largest values of $0.7\lambda_{i1}^{\mathrm{ref}}+0.3\lambda_{i2}^{\mathrm{ref}}$ are treated after period 20. The spatial components based on six modes and on the full graph use the same underlying Gaussian coefficients; the loading mixtures also share their unstructured draws.

The observations supplied to the estimators are $X_{it}=Z_{it}+\nu\epsilon_{it}$, with independent standard-normal $\epsilon_{it}$ and $\nu\in\{0.25,0.5,1,2,3,4,5\}$. We reuse the standardized errors across settings within a replication. Before fitting either probabilistic model, we center and scale the observations using their mean and sample standard deviation over $\Omega$. Both Gaussian models include an intercept and normalized Helmert contrasts for unit and period effects, and estimate a common observation variance. Their unit-by-period interactions have AR(1) temporal dependence. Across units, the nonspatial interaction is independent. The spatial interaction combines scaled structured and unstructured components under the original Besag--York--Molli\'e (BYM) parameterization with separate precision parameters \citep{besag1991bayesian}. We use this parameterization for computational efficiency. Experiment 2 and the spatial applications use BYM2, which expresses the structured and unstructured components through an overall scale and a spatial mixing parameter \citep{riebler2016intuitive}. We assign priors to the component scales here and directly to the overall scale and mixing parameter in the BYM2 analyses.

On the standardized scale, let $s_{\mathrm{res}}$ denote the residual standard deviation, with a residual degrees-of-freedom adjustment, from an additive unit-and-period-effects regression on $\Omega$. We set $U_{\mathrm{ST}}=2s_{\mathrm{res}}$ and calibrate the interaction-scale priors to satisfy
\[
\Pr(\sigma_{\mathrm{ST}}>U_{\mathrm{ST}})=0.05.
\]
For the nonspatial model, we use a penalized-complexity prior on $\sigma_{\mathrm{ST}}$. For the spatial model, write $\sigma_u$ and $\sigma_v$ for the unstructured and structured component scales and define $\sigma_{\mathrm{ST}}=(\sigma_u^2+\sigma_v^2)^{1/2}$. The two component penalized-complexity priors are calibrated jointly so that this combined interaction scale satisfies the same tail condition.

Fixed-effect coefficients have independent $N(0,2.5^2)$ priors on the standardized scale. The Gaussian observation precision has a shape--rate $\operatorname{Gamma}(1,0.00005)$ prior. For temporal correlation, we use the transformed parameter $\theta_\rho$, where $\rho=2\operatorname{logit}^{-1}(\theta_\rho)-1$, and assign $\theta_\rho\sim N(0,0.15^{-1})$. We obtain 100 joint posterior draws from each model fit and transform the complete latent panels back to the original scale before applying GSC.

For each observed or sampled latent panel, we remove the unit and period means calculated from the control panel and estimate two temporal factors by singular-value decomposition. We then regress each treated unit's pretreatment series on an intercept and the factor paths. Its fitted coefficients and the post-treatment factors determine the counterfactual imputations. For the probabilistic analyses, the point estimate averages these imputations across posterior draws. Rank is fixed at two throughout this experiment.

The main comparison is mean squared error (MSE) for the average latent counterfactual over the 100 treated post-treatment cells, relative to GSC fitted to observed outcomes. We also report error for individual missing cells and reconstruction error on the observed untreated cells. Reconstruction uses the observed outcomes or posterior means before applying the causal estimator. Pointwise 95\% Monte Carlo intervals use 2,000 paired bootstrap resamples of the 1,000 replications.

\subsubsection{Results}

When the factor loadings are fully spatially structured, the spatial model lowers the MSE of the average counterfactual at every examined noise level (Figure~\ref{fig:pcpa-risk} in the main paper). The nonspatial model has higher aggregate MSE throughout this range. With mixed loadings, the spatial model has lower aggregate MSE than GSC fitted to observed outcomes at $\nu=3,4,5$. With unstructured loadings, both probabilistic models have lower aggregate MSE only at $\nu=5$.

Improvements in reconstruction do not always extend to counterfactual imputation (Table~\ref{tab:pcpa-reconstruction}). With unstructured loadings and $\nu=1$, observed-cell reconstruction RMSE falls from 1.000 to 0.655 and 0.654 under the nonspatial and spatial models, respectively. Aggregate imputation RMSE nevertheless increases from 0.196 for GSC fitted to observed outcomes to 0.311 and 0.313. At $\nu=5$, aggregate imputation RMSE is instead 0.806 for GSC fitted to observed outcomes, compared with 0.791 and 0.781 for the two probabilistic analyses.

\begin{table}[H]
\sisetup{group-digits=integer}
\centering
\begin{threeparttable}
\caption{Reconstruction and counterfactual imputation accuracy.}
\label{tab:pcpa-reconstruction}
\label{tab:experiment1-three-risk}
\small
\setstretch{1}
\setlength{\tabcolsep}{0pt}
\begin{tabular*}{\linewidth}{@{}S[table-format=1.2,table-column-width=\dimexpr\linewidth-397pt\relax]
@{\hspace{8pt}}S[table-format=1.3,table-column-width=39pt]
@{\hspace{3pt}}S[table-format=1.3,table-column-width=39pt]
@{\hspace{3pt}}S[table-format=1.3,table-column-width=39pt]
@{\hspace{10pt}}S[table-format=1.3,table-column-width=39pt]
@{\hspace{3pt}}S[table-format=1.3,table-column-width=39pt]
@{\hspace{3pt}}S[table-format=1.3,table-column-width=39pt]
@{\hspace{10pt}}S[table-format=1.3,table-column-width=39pt]
@{\hspace{3pt}}S[table-format=1.3,table-column-width=39pt]
@{\hspace{3pt}}S[table-format=1.3,table-column-width=39pt]@{}}
\toprule
& \multicolumn{3}{c}{\shortstack{Observed-cell\\ reconstruction}} & \multicolumn{3}{c}{\shortstack{Missing-cell\\ imputation}} & \multicolumn{3}{c}{\shortstack{Aggregate\\ imputation}} \\
\cmidrule(lr){2-4} \cmidrule(lr){5-7} \cmidrule(lr){8-10}
{$\nu$} & {Raw} & {Nonsp.} & {Spatial} & {GSC} & {Nonsp.} & {Spatial} & {GSC} & {Nonsp.} & {Spatial} \\
\midrule
\multicolumn{10}{l}{\textit{Loading share $\gamma=0$}} \\
0.25 & 0.250 & 0.261 & 0.261 & 0.158 & 0.257 & 0.258 & 0.046 & 0.068 & 0.068 \\
0.5 & 0.500 & 0.431 & 0.429 & 0.317 & 0.451 & 0.454 & 0.093 & 0.139 & 0.141 \\
1 & 1.000 & 0.655 & 0.654 & 0.635 & 0.810 & 0.815 & 0.196 & 0.311 & 0.313 \\
2 & 1.999 & 1.126 & 1.044 & 1.261 & 1.232 & 1.236 & 0.432 & 0.533 & 0.544 \\
3 & 2.999 & 1.834 & 1.452 & 1.761 & 1.518 & 1.481 & 0.602 & 0.635 & 0.649 \\
4 & 3.999 & 2.726 & 1.829 & 2.191 & 1.854 & 1.686 & 0.714 & 0.718 & 0.717 \\
5 & 4.998 & 3.650 & 2.286 & 2.602 & 2.215 & 1.939 & 0.806 & 0.791 & 0.781 \\
\addlinespace
\multicolumn{10}{l}{\textit{Loading share $\gamma=0.5$}} \\
0.25 & 0.250 & 0.261 & 0.235 & 0.164 & 0.278 & 0.232 & 0.050 & 0.078 & 0.074 \\
0.5 & 0.500 & 0.426 & 0.376 & 0.328 & 0.485 & 0.440 & 0.102 & 0.163 & 0.153 \\
1 & 1.000 & 0.657 & 0.553 & 0.661 & 0.859 & 0.780 & 0.217 & 0.362 & 0.323 \\
2 & 1.999 & 1.117 & 0.846 & 1.307 & 1.296 & 1.160 & 0.492 & 0.633 & 0.536 \\
3 & 2.999 & 1.824 & 1.194 & 1.808 & 1.590 & 1.416 & 0.675 & 0.742 & 0.654 \\
4 & 3.999 & 2.744 & 1.651 & 2.235 & 1.922 & 1.655 & 0.796 & 0.817 & 0.740 \\
5 & 4.998 & 3.696 & 2.126 & 2.649 & 2.268 & 1.912 & 0.889 & 0.881 & 0.820 \\
\addlinespace
\multicolumn{10}{l}{\textit{Loading share $\gamma=1$}} \\
0.25 & 0.250 & 0.262 & 0.237 & 0.154 & 0.252 & 0.149 & 0.042 & 0.058 & 0.041 \\
0.5 & 0.500 & 0.423 & 0.339 & 0.306 & 0.426 & 0.258 & 0.084 & 0.125 & 0.076 \\
1 & 1.000 & 0.665 & 0.456 & 0.608 & 0.730 & 0.456 & 0.173 & 0.257 & 0.143 \\
2 & 1.999 & 1.158 & 0.720 & 1.164 & 1.125 & 0.805 & 0.335 & 0.445 & 0.266 \\
3 & 2.999 & 1.889 & 1.082 & 1.653 & 1.447 & 1.116 & 0.473 & 0.550 & 0.384 \\
4 & 3.999 & 2.764 & 1.591 & 2.103 & 1.787 & 1.432 & 0.592 & 0.646 & 0.492 \\
5 & 4.998 & 3.627 & 2.055 & 2.536 & 2.139 & 1.720 & 0.700 & 0.720 & 0.588 \\
\bottomrule
\end{tabular*}
\begin{tablenotes}[flushleft]
\footnotesize
\item Notes: Root mean squared errors over 1,000 replications. Reconstruction uses observed outcomes (Raw) or posterior means on observed untreated cells; imputation uses rank-two GSC fitted to observed or sampled latent panels. Aggregate error is calculated after averaging over the 100 treated cells. $\nu$ is the generating observation-error standard deviation and $\gamma$ the spatially structured share of the factor loadings. Nonsp. denotes the nonspatial Gaussian model, and Spatial the Gaussian model with Besag--York--Molli\'e spatial structure. Both estimate a common observation variance and have AR(1) temporal dependence. The main figure reports MSE differences for the aggregate comparison.
\end{tablenotes}
\end{threeparttable}
\end{table}

\FloatBarrier

\subsection{Predictive inference for aggregate effects}
\label{app:experiment2-design}
\label{subsec:pcpa-experiment-prediction}
\label{subsec:experiment-prediction}

Imputing the systematic outcome process leaves uncertainty about the realized untreated outcomes. For average treatment effects, the dependence of that remaining variation matters as well as its magnitude. We vary both in the second experiment and assess predictive inference for the realized effect. Each replication includes rank selection and estimation of the predictive mixture weights.

\subsubsection{Design and implementation}

We distinguish the systematic component $M_{it}$ from the additional latent variation introduced below. The systematic component follows the additive rank-two construction in Appendix~\ref{app:experiment1-design}, with $N=60$, $T=30$, and $T_0=20$. The unit effects, period effects, factor processes, and burn-in are the same as in the first experiment. For the loadings, the structured component now uses the full scaled Besag covariance and contributes a share of $0.5$. We sample ten treated units without replacement with probabilities proportional to $\exp(0.8\lambda_{i1}+0.4\lambda_{i2})$.

We add latent innovations through $Z_{it}(0)=M_{it}+U_{it}$, where
\[
U_{it}
=
\sigma_U
\left\{\sqrt{1-\psi}\,\xi_{it}+\sqrt{\psi}\,G_{it}\right\}.
\]
The $\xi_{it}$ are independent standard normal. Independently, $\vect G_t$ follows a stationary AR(1) process with correlation $0.65$ and marginal spatial covariance $0.4\mat I+0.6\mat K_B$, where $\mat K_B$ is the constrained Besag covariance scaled to unit average marginal variance. The settings labeled None, Independent, Mixed, Correlated, and Correlated with doubled scale use
\[
(\sigma_U,\psi)
=
(0,0),\ (0.35,0),\ (0.35,0.5),\ (0.35,1),\ (0.70,1),
\]
respectively. Within a replication, all regimes share the systematic component and standardized innovation draws.

Untreated observations are Gaussian with means $Z_{it}(0)$ and variances $1/e_{it}$. To generate heterogeneous precisions, we draw lognormal multipliers truncated to $[0.25,4]$, then divide them by their panel mean. Before truncation, the log-scale mean is minus one half the log-scale variance; we choose the log-scale standard deviation numerically so that the truncated distribution has coefficient of variation $0.8$. The fitted likelihood uses $e_{it}$ as known relative precisions with common precision one. Treatment adds $0.50$ to the realized untreated outcome in each treated cell, giving a realized sample average treatment effect on the treated (SATT) of $0.50$.

We fit a Gaussian model with a BYM2 interaction \citep{riebler2016intuitive} and AR(1) temporal dependence. As in the spatial empirical models, we specify priors on the overall interaction scale and spatial mixing parameter. Normalized Helmert fixed-effect coefficients have independent $\N(0,2.5^2)$ priors. The transformed temporal-correlation parameter $\theta_\rho$, defined in Appendix~\ref{app:experiment1-design}, has a normal prior with mean zero and precision $0.15$. The spatial component is scaled and constrained to sum to zero, with penalized-complexity priors satisfying $\Pr(\sigma>1)=0.01$ and $\Pr(\phi<0.5)=0.5$, where $\sigma$ is the interaction scale and $\phi$ its spatial mixing parameter. Reference innovations are sampled from the fitted conditional covariance as described in Appendix~\ref{app:pcpa-reference-sampling}.

The validation blocks match the ten-unit, ten-period treatment pattern. In each replication, we make two balanced random partitions of the fifty controls, giving ten folds in total. Each fold withholds ten controls over the ten post-treatment periods, leaving forty fully observed controls. Using 256 draws from the approximate posterior, we select the GSC rank from zero through four by Equation~\eqref{eq:pcpa-point-score}. We then hold the selected rank fixed and weight the folds equally when calibrating the predictive distribution of the SATT.

For final prediction, we use 512 joint posterior draws and the five candidate innovation distributions in Appendix~\ref{app:pcpa-innovation-library}. We compare calibrated prediction with prediction without added latent innovations. Both use the same fitted probabilistic model and selected GSC rank, and both include posterior uncertainty and observation variation. Each reference innovation is paired with the model parameters from the same posterior draw. Observation noise is shared across candidates where possible.

To examine prediction at other aggregation levels, we use the same mixture weights to evaluate coverage for the ten event-time averages and the 100 individual treated post-treatment cells. We report nominal levels of 50\%, 80\%, 90\%, and 95\%, without recalibrating the weights. We use 500 replications per regime. For a mean paired CRPS difference $\bar d$, the Monte Carlo interval is $\bar d\pm1.96s_d/\sqrt{500}$, where $s_d$ is the standard deviation of the replication-level differences.

\subsubsection{Results}

Allowing latent innovations brings SATT coverage closer to 95\% in every regime (Table~\ref{tab:pcpa-predictive-performance-full}). The largest gain occurs with correlated innovations at the larger scale, where coverage increases from 88.4\% to 93.8\%. The corresponding CRPS difference, multiplied by 1,000, is $-2.1$ with a paired 95\% Monte Carlo interval of $[-3.7,-0.6]$. In the other four regimes, coverage increases by 2.0--2.8 percentage points, while mean CRPS changes little. The paired intervals for these smaller score differences include zero.

\begin{table}[!tbp]
\centering
\begin{threeparttable}
\caption{Predictive scores and coverage for the SATT.}
\label{tab:pcpa-predictive-performance-full}
\label{tab:experiment2-stacking}
\small\setstretch{1}
\setlength{\tabcolsep}{0pt}
\begin{tabular*}{\linewidth}{@{}>{\raggedright\arraybackslash}p{\dimexpr\linewidth-362pt\relax}
@{\hspace{6pt}}S[table-format=3.1,table-column-width=40pt]
@{\hspace{5pt}}S[table-format=2.1,table-column-width=48pt]
@{\hspace{12pt}}S[table-format=3.1,table-column-width=40pt]
@{\hspace{5pt}}S[table-format=2.1,table-column-width=48pt]
@{\hspace{12pt}}S[table-format=-1.1,table-column-width=65pt]
@{\hspace{4pt}}>{\centering\arraybackslash}p{77pt}@{}}
\toprule
& \multicolumn{2}{c}{\shortstack[c]{No added latent\\innovations}}
& \multicolumn{2}{c}{\shortstack[c]{Calibrated\\prediction}}
& \multicolumn{2}{c}{CRPS difference} \\
\cmidrule(lr){2-3}\cmidrule(lr){4-5}\cmidrule(lr){6-7}
Regime & {CRPS} & {Coverage} & {CRPS} & {Coverage} & {Estimate} & 95\% MC interval \\
 & {\(\times10^3\)} & {(\%)} & {\(\times10^3\)} & {(\%)} & {\(\times10^3\)} & \(\times10^3\) \\
\midrule
No innovation & 153.5 & 92.8 & 153.0 & 94.8 & -0.5 & $[-1.3,\,0.3]$ \\
Independent & 155.5 & 91.8 & 155.0 & 94.4 & -0.5 & $[-1.3,\,0.4]$ \\
Mixed & 158.3 & 92.0 & 157.8 & 94.2 & -0.5 & $[-1.4,\,0.4]$ \\
Correlated & 162.8 & 90.4 & 162.1 & 93.2 & -0.6 & $[-1.5,\,0.3]$ \\
Correlated, doubled scale & 195.3 & 88.4 & 193.2 & 93.8 & -2.1 & $[-3.7,\,-0.6]$ \\
\bottomrule
\end{tabular*}
\begin{tablenotes}[flushleft]
\footnotesize
\item Notes: Mean continuous ranked probability scores (CRPS) and coverage of 95\% predictive intervals over 500 replications per setting, with a realized average treatment effect of 0.50. Both predictions use the same fitted probabilistic model and cross-validated GSC rank, and include posterior uncertainty and observation variation. Calibrated prediction additionally allows latent innovations, with mixture weights estimated by predictive stacking on held-out untreated outcomes. The final columns report differences in CRPS between calibrated prediction and prediction without added latent innovations, in that order, with paired $95\%$ Monte Carlo intervals. Lower CRPS is better.
\end{tablenotes}
\end{threeparttable}
\end{table}

With weights calibrated for the SATT, coverage is also closer to 95\% for event-time averages and individual treated cells (Table~\ref{tab:pcpa-coverage-levels}). It ranges from 92.3\% to 94.0\% for event-time averages and from 95.6\% to 95.9\% for individual cells. These results use the original weights without further calibration.

\begin{table}[!tbp]
\centering
\begin{threeparttable}
\caption{Coverage of predictive intervals across aggregation levels in Experiment 2.}
\label{tab:pcpa-coverage-levels}
\label{tab:experiment2-secondary-coverage}
\small\setstretch{1}
\setlength{\tabcolsep}{0pt}
\begin{tabular*}{\linewidth}{@{}>{\raggedright\arraybackslash}p{\dimexpr\linewidth-297pt\relax}
@{\hspace{5pt}}S[table-format=2.1,table-column-width=32pt]
@{\hspace{4pt}}S[table-format=2.1,table-column-width=32pt]
@{\hspace{4pt}}S[table-format=2.1,table-column-width=32pt]
@{\hspace{4pt}}S[table-format=2.1,table-column-width=32pt]
@{\hspace{12pt}}S[table-format=2.1,table-column-width=32pt]
@{\hspace{4pt}}S[table-format=2.1,table-column-width=32pt]
@{\hspace{4pt}}S[table-format=2.1,table-column-width=32pt]
@{\hspace{4pt}}S[table-format=2.1,table-column-width=32pt]
@{}}
\toprule
& \multicolumn{4}{c}{No added latent innovations}
& \multicolumn{4}{c}{Calibrated prediction} \\
\cmidrule(lr){2-5}\cmidrule(lr){6-9}
Regime & {50\%} & {80\%} & {90\%} & {95\%} & {50\%} & {80\%} & {90\%} & {95\%} \\
\midrule
\multicolumn{9}{@{}l}{\textit{Panel A: SATT}} \\
\addlinespace[2pt]
No innovation & 42.2 & 72.8 & 86.0 & 92.8 & 45.4 & 76.4 & 89.0 & 94.8 \\
Independent & 44.4 & 71.6 & 85.2 & 91.8 & 48.6 & 78.2 & 89.6 & 94.4 \\
Mixed & 43.8 & 74.2 & 84.4 & 92.0 & 48.0 & 78.4 & 89.0 & 94.2 \\
Correlated & 44.4 & 75.6 & 85.0 & 90.4 & 47.6 & 78.6 & 89.0 & 93.2 \\
Correlated, doubled scale & 41.0 & 71.4 & 82.2 & 88.4 & 45.8 & 79.0 & 89.0 & 93.8 \\
\addlinespace[5pt]
\multicolumn{9}{@{}l}{\textit{Panel B: Event-time averages}} \\
\addlinespace[2pt]
No innovation & 43.0 & 72.4 & 83.1 & 89.6 & 46.6 & 75.8 & 86.2 & 92.3 \\
Independent & 42.4 & 71.8 & 83.3 & 89.9 & 47.1 & 76.2 & 87.0 & 93.0 \\
Mixed & 42.5 & 71.8 & 83.3 & 90.2 & 46.9 & 76.6 & 87.0 & 93.0 \\
Correlated & 43.2 & 71.6 & 83.5 & 90.2 & 47.2 & 76.5 & 87.6 & 93.0 \\
Correlated, doubled scale & 41.3 & 70.6 & 82.7 & 89.3 & 47.7 & 78.1 & 88.5 & 94.0 \\
\addlinespace[5pt]
\multicolumn{9}{@{}l}{\textit{Panel C: Individual treated cells}} \\
\addlinespace[2pt]
No innovation & 48.0 & 77.9 & 88.3 & 93.9 & 52.6 & 82.4 & 91.4 & 95.7 \\
Independent & 46.7 & 76.6 & 87.3 & 93.1 & 52.7 & 82.1 & 91.1 & 95.6 \\
Mixed & 47.4 & 76.9 & 87.8 & 93.4 & 52.8 & 82.2 & 91.4 & 95.7 \\
Correlated & 48.0 & 77.4 & 88.0 & 93.6 & 53.5 & 82.6 & 91.6 & 95.9 \\
Correlated, doubled scale & 46.0 & 74.4 & 85.5 & 91.6 & 53.9 & 82.6 & 91.7 & 95.8 \\
\bottomrule
\end{tabular*}
\begin{tablenotes}[flushleft]
\footnotesize
\item Notes: Coverage of central predictive intervals for true effects of 0.50, evaluated over 500 replications per regime at the nominal levels shown. In Panels B and C, coverage is averaged over the ten event-time estimates and 100 treated cells, respectively. Mixture weights calibrated for the SATT are used at all three aggregation levels without retuning. The selected GSC rank is shared by the predictions with and without added latent innovations.
\end{tablenotes}
\end{threeparttable}
\end{table}

\FloatBarrier

\subsection{Poisson comparative-case studies}
\label{app:experiment3-design}
\label{subsec:pcpa-experiment-poisson}
\label{subsec:experiment-comparative}

The third experiment examines non-Gaussian outcomes in comparative case studies. We compare synthetic controls (SCM) and ridge-augmented synthetic controls (ASCM; \citealp{ben2021augmented}) using simulated Poisson counts. Comparisons with observed rates and empirical log rates separate the contribution of the observation model from that of changing the estimation scale.

\subsubsection{Design and implementation}

We generate twenty control units and two treated cases over 35 periods, with 30 pretreatment periods. Their latent untreated log rates follow
\[
Z_{it}(0)=\mu+\alpha_i+\delta_t+0.35\lambda_i f_t.
\]
We generate the control intercepts and loadings from separate sets of independent standard-normal draws. Each set is centered across the twenty controls and multiplied by $\sqrt{20/19}$ to give unit marginal variance. Multiplying the intercepts by $0.25$ then gives their specified marginal standard deviation. The factor follows a stationary AR(1) process with correlation $0.8$ and unit marginal variance. The period effect has correlation $0.6$ and marginal standard deviation $0.1$. We discard 100 initial periods from both paths.

To examine both interpolation and extrapolation, we place one treated case inside and the other outside the convex hull of the control coefficient pairs $(\alpha_i,\lambda_i)$. Let $\vect c$ be the centroid of those control pairs and $\vect m_4$ the average of four randomly selected pairs. The interpolation case has coefficients $0.70\vect c+0.30\vect m_4$. For the extrapolation case, we choose the convex-hull edge whose midpoint distance from $\vect c$ is closest to the median of these distances. Writing its midpoint as $\vect m_e$, the treated coefficients are $\vect m_e+0.10(\vect m_e-\vect c)$. They lie outside the convex hull but have the affine representation $1.10\vect m_e-0.10\vect c$. The treated cases do not serve as controls for one another.

We choose the common intercept so that the mean expected count in the control pretreatment panel is eight. With exposure $E_{it}=8$, we generate
\[
Y_{it}(0)\sim\operatorname{Poisson}\{8\exp(Z_{it}(0))\}.
\]
We add no further unit-period latent innovations. Treatment adds two events to each realized untreated post-treatment count, fixing the realized rate effect at $0.25$.

We fit one Poisson model with an intercept, constrained unit and period effects, and a unit-by-period interaction with AR(1) temporal dependence that is independent across units. Both treated post-treatment trajectories are excluded from the likelihood. Coefficient priors are $\N(0,2.5^2)$, the interaction precision has a shape--rate $\operatorname{Gamma}(1,0.00005)$ prior, and the transformed temporal-correlation parameter $\theta_\rho$ has a normal prior with mean zero and precision $0.15$, using the transformation in Appendix~\ref{app:experiment1-design}. Each of 500 replications uses 500 joint posterior draws and 256 draws from the approximate posterior per validation fold.

We use the synthetic-control and ridge-augmented estimators described in Appendix~\ref{app:pcpa-estimators}. The augmentation penalty ratio is fixed at $r=1$ for simplicity; the penalty itself is recalculated for each input panel using the scaling rule in Appendix~\ref{app:pcpa-ascm-implementation}.

To assess point estimation, we apply each estimator to observed rates, empirical log rates $\log\{(Y+0.5)/8\}$, and posterior draws of latent log rates. For each treated case, we evaluate accuracy for the average untreated mean rate,
\[
\frac{1}{5}\sum_{t>T_0}\exp\{Z_{it}(0)\}.
\]
For each posterior draw, we calculate $\exp\{c_{k,it}(\mathbf Z_\Omega^{(s)}(0))\}$ and average these imputed mean rates over periods and draws.

For predictive calibration, blocks begin at periods $10,12,\ldots,26$ within each case's pretreatment history. We withhold that case's outcomes from the start onward and evaluate predictions over the next five periods, using all twenty controls. We use the resulting eighteen approximate posteriors for both estimators and estimate the predictive mixture weights separately for each estimator and treated case.

For each joint posterior draw, we apply $c_k\{\mathbf Z_\Omega^{(s)}(0)\}$ and generate untreated counts through the Poisson observation model, evaluating the reference innovation covariance at $\boldsymbol\theta^{(s)}$. We evaluate the predictive distribution of the five-period average realized rate effect at $0.25$, using the same 500 replications as the point comparison.

\subsubsection{Results}

In both simulated cases, SCM and ASCM applied to posterior draws of latent log rates have lower RMSE than when applied to empirical log rates (Table~\ref{tab:pcpa-poisson}, Panel A). Panel A also reports the results from observed rates.

For predictive inference, coverage of the 95\% intervals from the calibrated mixture ranges from 93.0\% to 98.0\% across the four settings (Panel B). The corresponding changes in mean CRPS are small.

\begin{table}[!tbp]
\centering
\begin{threeparttable}
\caption{Imputation accuracy and predictive inference in the Poisson experiment.}
\label{tab:pcpa-poisson}
\label{tab:experiment3-comparative}
\small\setstretch{1}
\setlength{\tabcolsep}{0pt}
\begin{tabular*}{\linewidth}{@{}>{\raggedright\arraybackslash}p{0.24\linewidth}
@{\extracolsep{\fill}}>{\raggedright\arraybackslash}p{0.18\linewidth}
S[table-format=1.3]S[table-format=1.3]S[table-format=1.3]@{}}
\toprule
\multicolumn{5}{l}{\textit{Panel A. RMSE for the average untreated mean rate}} \\
\addlinespace[3pt]
Design & Estimator & {Observed rate} & {Empirical log rate} & {Posterior latent log rates} \\
\midrule
\multicolumn{5}{l}{\textit{Interpolation}} \\
 & SCM  & 0.118 & 0.113 & 0.082 \\
 & ASCM & 0.141 & 0.134 & 0.089 \\
\addlinespace[4pt]
\multicolumn{5}{l}{\textit{Extrapolation}} \\
 & SCM  & 0.187 & 0.247 & 0.230 \\
 & ASCM & 0.172 & 0.200 & 0.189 \\
\bottomrule
\end{tabular*}
\par\medskip
\begin{tabular*}{\linewidth}{@{}>{\raggedright\arraybackslash}p{0.20\linewidth}
>{\raggedright\arraybackslash}p{0.15\linewidth}
@{\extracolsep{\fill}}S[table-format=1.3]S[table-format=2.1]
S[table-format=1.3]S[table-format=2.1]@{}}
\multicolumn{6}{l}{\textit{Panel B. Predictive inference for the realized rate effect}} \\
\addlinespace[3pt]
& & \multicolumn{2}{c}{No added latent innovations}
& \multicolumn{2}{c}{Calibrated prediction} \\
\cmidrule(lr){3-4}\cmidrule(lr){5-6}
Design & Estimator & {CRPS} & {Coverage (\%)} & {CRPS} & {Coverage (\%)} \\
\midrule
\multicolumn{6}{l}{\textit{Interpolation}} \\
 & SCM  & 0.095 & 97.0 & 0.097 & 98.0 \\
 & ASCM & 0.097 & 97.4 & 0.099 & 98.0 \\
\addlinespace[4pt]
\multicolumn{6}{l}{\textit{Extrapolation}} \\
 & SCM  & 0.155 & 85.6 & 0.152 & 93.0 \\
 & ASCM & 0.137 & 92.2 & 0.142 & 95.0 \\
\bottomrule
\end{tabular*}
\begin{tablenotes}[flushleft]
\footnotesize
\item Notes: Results for SCM and ASCM over 500 replications. Panel A reports RMSE for the five-period average untreated mean rate. The empirical log rate is $\log\{(Y+0.5)/8\}$. Panel B reports CRPS and coverage of 95\% predictive intervals for the five-period average realized rate effect of 0.25. Both predictive specifications include Poisson observation variation; calibrated prediction permits added latent innovations. The augmentation penalty ratio is fixed at one. Predictive mixture weights are estimated separately for each treated case and estimator.
\end{tablenotes}
\end{threeparttable}
\end{table}

\FloatBarrier

\section{Empirical examples: data, methods, and additional results}
\label{app:pcpa-empirical}
\label{app:pcpa-applications}
\label{app:empirical-details}

This appendix gives the full policy evaluations summarized in Section~\ref{subsec:pcpa-empirical-illustrations}. We describe the probabilistic models used in the applications, followed by the data construction, validation design, and results for each evaluation. The final subsection examines sensitivity to the candidate innovation scales and to influential validation holdouts.

\subsection{Model specification and computation}
\label{app:pcpa-empirical-shared}
\label{app:empirical-causal}

The probabilistic models used in the primary empirical analyses include an intercept, unit and period effects, and a unit-by-period interaction. We use a Gaussian observation model for cigarette sales and Poisson observation models for opioid deaths and traffic fatalities, with population and passenger-kilometers as exposures. The latent outcomes are sales in the first application and log rates in the two count-data applications.

The interaction allows units to follow different trajectories after accounting for additive unit and period effects. In the nonspatial specification, these trajectories are independent across units and follow a first-order autoregressive process over time. The spatial specification combines graph-structured and independent variation using the scaled Besag--York--Molli\'e formulation of \citet{riebler2016intuitive}. For the vector of interactions in period $t$, write
\[
\mathbf b_t
=\tau^{-1/2}
 \left\{\sqrt{\phi}\,\mathbf u_t^\star+
         \sqrt{1-\phi}\,\mathbf v_t\right\},
\qquad 0<\phi<1,
\]
where $\mathbf u_t^\star$ is the scaled Besag field defined by the adjacency graph and $\mathbf v_t$ has independent standard-normal entries across units. The two components are independent Gaussian processes and have the same first-order autoregressive correlation $\rho$ over time, giving a separable spatiotemporal covariance. The parameter $\tau^{-1/2}$ controls the interaction scale, and $\phi$ the contribution of the structured component.

Conditional on the parameters, the interaction approaches independence across units as $\phi$ approaches zero; setting $\rho=0$ removes its temporal dependence. We scale the structured component and constrain its values to sum to zero within each connected component of the graph, handling singleton components separately. For each spatial specification, we restrict the common adjacency graph to the included jurisdictions. In the opioid-mortality panel, Hawaii is the only jurisdiction with no neighbors in this graph and is assigned a separate precision parameter. The nonspatial models estimate their own interaction precision and temporal correlation.

Fixed-effect coefficients have independent $N(0,2.5^2)$ priors. Within each application, we use the same penalized-complexity prior \citep{fuglstad2019constructing} for the overall interaction standard deviation in the nonspatial and spatial specifications. Writing $\sigma_{\mathrm{ST}}=\tau^{-1/2}$ for this scale, we set
\[
\Pr(\sigma_{\mathrm{ST}}>U_{\mathrm{ST}})=0.05.
\]
For cigarette sales, we standardize using the mean and sample standard deviation over $\Omega$ and set $U_{\mathrm{ST}}=2s_{\mathrm{res}}$, where $s_{\mathrm{res}}$ is the residual standard deviation, adjusted for residual degrees of freedom, from a unit-and-period-effects regression of the standardized sales on $\Omega$. This gives $U_{\mathrm{ST}}=0.7309$. During validation, we retain the standardization and prior scale calculated from all untreated observations. For the two Poisson applications, $U_{\mathrm{ST}}=\log(2)$ on the log-rate scale.

The Gaussian observation precision has a shape--rate $\operatorname{Gamma}(1,0.00005)$ prior. For the spatial mixing parameter, we use $\Pr(\phi<0.5)=0.5$. For temporal correlation, we use $\theta_\rho\sim N(0,0.15^{-1})$, with $\rho=2\operatorname{logit}^{-1}(\theta_\rho)-1$.

We obtain 16,000 joint posterior draws for each cigarette-sales and opioid-mortality model. For Vision Zero, we use 32,000 draws to reduce the variability across random seeds observed in preliminary runs with 16,000 draws. In each validation fold, we use 512 draws from the approximate posterior and 2,048 predictive draws per candidate. The Poisson models are fitted using R-INLA's simplified Laplace approximation. Model and estimator selection and predictive calibration use the criteria in Appendix~\ref{app:pcpa-predictive}, with the holdouts described for each application below. Table~\ref{tab:pcpa-empirical-hyperparameters} reports posterior summaries of temporal correlation and the spatial share.

\begin{table}[!tbp]
\centering
\begin{minipage}{\linewidth}
\begin{threeparttable}
\caption{Estimated temporal and spatial dependence in the empirical models.}
\label{tab:pcpa-empirical-hyperparameters}
\label{tab:pcpa-empirical-dependence}
\label{tab:pcpa-hyperparameters}
\label{tab:empirical-hyperparameters}
\small\setstretch{1}
\setlength{\tabcolsep}{0pt}
\begin{tabular*}{\linewidth}{@{}>{\raggedright\arraybackslash}p{\dimexpr\linewidth-318pt\relax}
@{\hspace{10pt}}>{\raggedright\arraybackslash}p{81pt}
@{\hspace{10pt}}>{\centering\arraybackslash}p{108pt}
@{\hspace{10pt}}>{\centering\arraybackslash}p{99pt}@{}}
\toprule
Application & Probabilistic model & AR(1) correlation & Spatial share \\
\midrule
California & Nonspatial & 0.981 [0.959, 0.991] & -- \\
California & Spatial & 0.980 [0.960, 0.991] & 0.06 [0.02, 0.15] \\
Cannabis & Nonspatial & 0.966 [0.933, 0.985] & -- \\
Cannabis & Spatial & 0.949 [0.903, 0.976] & 0.38 [0.25, 0.54] \\
Vision Zero & Nonspatial & 0.992 [0.978, 0.998] & -- \\
\bottomrule
\end{tabular*}
\begin{tablenotes}[flushleft]
\footnotesize
\item Notes: Posterior medians with 95\% credible intervals. The temporal correlation $\rho$ describes autoregressive dependence in the unit-by-period interaction. The spatial share $\phi$ measures the contribution of its graph-structured component, as defined in Appendix~\ref{app:pcpa-empirical-shared}. Nonspatial models have no spatial component; dashes indicate that no spatial mixing parameter is estimated.
\end{tablenotes}
\end{threeparttable}
\end{minipage}
\end{table}

\subsection{California's tobacco-control program (Proposition 99)}
\label{subsec:pcpa-california}
\label{subsec:california}

We revisit the evaluation of California's tobacco-control program by \citet{abadie2010synthetic}.

\subsubsection{Data and analysis}

The panel contains California and 38 controls observed from 1970 through 2000. The outcome is cigarette sales per capita, and 1989 is the first post-intervention year. We compare SCM fitted to observed sales with SCM combined with the nonspatial and spatial Gaussian models described above. We standardize sales as described in Appendix~\ref{app:pcpa-empirical-shared} and transform the latent panel draws back to packs per capita before applying SCM, expressing the reference innovations on the same scale.

We use the original matching predictors in all synthetic-control fits: mean log income, retail price, and the share aged 15--24 over 1980--1988; mean beer consumption over 1984--1988; and cigarette sales in 1975, 1980, and 1988. For each probabilistic specification and validation fold, we determine the predictor-weighting matrix from posterior means of the latent panel and hold it fixed while re-estimating control weights across draws.

The validation design matches California's twelve-year post-intervention period. We withhold each of the 38 control-state trajectories over 1989--2000, compare point predictions for individual state-years, and calibrate predictive distributions for the twelve-year average observed-minus-predicted contrast. Reported effects average over 1989--2000 and are expressed in packs per capita.

For the placebo comparisons, we assign the 1989 intervention date to each control state in turn and exclude that state and California from its control pool. We include placebo states whose pretreatment root mean squared prediction error is no more than five times California's \citep{abadie2010synthetic}.

\subsubsection{Results}

The estimated counterfactual trajectories are similar across the three analyses. The average effect is $-18.8$ packs per capita with SCM on observed sales. The nonspatial and spatial Gaussian analyses give predictive medians of $-18.8$ and $-18.7$, with $95\%$ predictive intervals of $[-42.5,5.3]$ and $[-40.7,3.9]$, respectively.

\begin{figure}[!tbp]
\centering
\caption{California's tobacco-control program.}
\includegraphics[width=\textwidth]{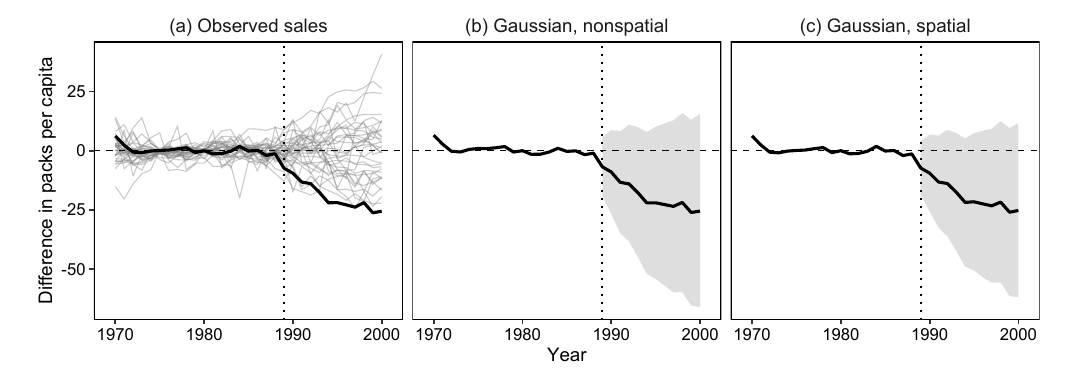}
\figurenote{Differences between observed and counterfactual cigarette sales, in packs per capita, using SCM with (a) observed sales, (b) the nonspatial Gaussian model, and (c) the spatial Gaussian model. Solid black lines are California's estimated differences; vertical dotted lines mark the 1989 intervention. Thin gray lines in (a) are placebo estimates for control states with pretreatment root mean squared prediction error no more than five times California's \citep{abadie2010synthetic}. Shaded bands in (b)--(c) are 95\% pointwise predictive intervals.}
\label{fig:pcpa-california}
\label{fig:pcpa-california-effect}
\label{fig:california}
\end{figure}

\begin{figure}[!tbp]
\centering
\caption{California's observed and counterfactual cigarette sales.}
\includegraphics[width=\textwidth]{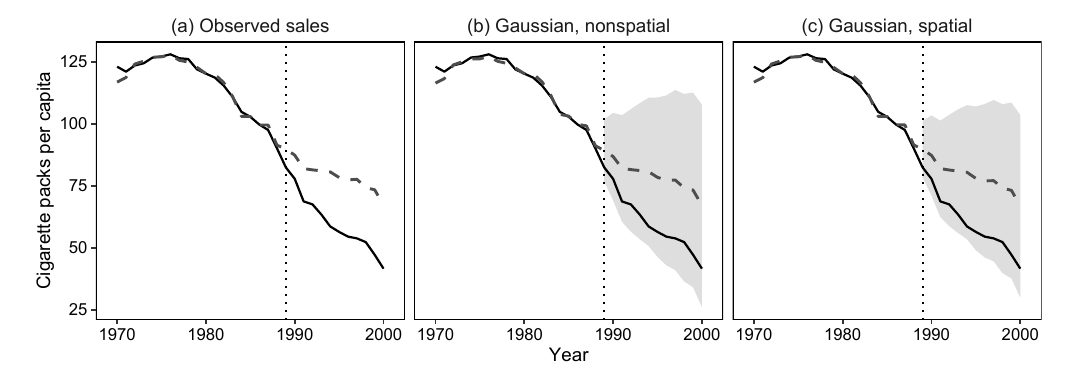}
\figurenote{Solid lines show observed sales and dashed lines the counterfactuals from SCM using (a) observed sales, (b) the nonspatial Gaussian model, and (c) the spatial Gaussian model. Rates are packs per capita; vertical dotted lines mark the 1989 intervention. Shaded bands in (b)--(c) are 95\% pointwise predictive intervals for counterfactual sales.}
\label{fig:pcpa-california-trajectories}
\label{fig:pcpa-california-counterfactual}
\end{figure}

Cross-validated RMSE is $12.44$ for observed sales, $13.20$ for the nonspatial model, and $12.24$ for the spatial model (Table~\ref{tab:pcpa-empirical-summary}). The spatial specification therefore performs best by this criterion, with a small improvement over SCM on observed sales. The posterior median spatial share is $0.06$, with $95\%$ credible interval $[0.02,0.15]$ (Table~\ref{tab:pcpa-empirical-dependence}). Figures~\ref{fig:pcpa-california-effect} and~\ref{fig:pcpa-california-counterfactual} show the estimated effects and counterfactual trajectories.

\FloatBarrier

\subsection{Recreational cannabis laws and opioid mortality}
\label{subsec:pcpa-cannabis}
\label{subsec:marijuana}

Evaluations of cannabis policies have reported mixed findings for opioid mortality \citep{chan2020effects,nguyen2024cannabis}. We revisit this setting to examine whether a Poisson observation model and spatial pooling improve counterfactual prediction. Geographic variation in mortality trends provides a reason to consider spatial dependence \citep{rigg2018opioid}.

\subsubsection{Data and analysis}

The panel contains opioid death counts and population exposures for 46 U.S. jurisdictions from 1999 through 2019, with six staggered adopters and policy timing as in \citet{nguyen2024cannabis}. We construct counts from CDC WONDER using underlying-cause codes X40--X44, X60--X64, X85, and Y10--Y14, together with opioid involvement recorded under T40.0--T40.4 or T40.6. The adopting jurisdictions are Colorado, Washington, Oregon, California, Nevada, and Massachusetts.

The models include four binary time-varying policy indicators: medical-cannabis dispensary availability, a Good Samaritan law, a naloxone-access law, and a mandatory prescription-drug-monitoring-program law \citep{nguyen2024cannabis}. We assign policy and dispensary dates in January--June to that calendar year and dates in July--December to the following year. The rule applies to recreational and medical dispensary availability and the other policy indicators. The four covariates are observed throughout the panel and enter both probabilistic specifications.

We compare GSC and matrix completion on observed mortality rates and on latent log rates from Poisson models with and without spatial pooling. Effects are expressed as deaths per 100,000 and averaged equally over the twenty treated state-years. For GSC, we consider factor ranks $R\in\{0,1,2,3,4\}$. For matrix completion, we evaluate penalty ratios
\[
\{0.01,0.03,0.10,0.20,0.30,0.50,0.70,1.00\}.
\]
For each input panel and holdout, let $\mathbf R$ contain the residuals after fitting the additive effects and policy covariates on the available untreated cells, with zeros elsewhere. Writing $m$ for the number of those cells, we set
\[
\lambda=r\lambda_{\max},
\qquad
\lambda_{\max}=\frac{2}{m}\|\mathbf R\|_{\mathrm{op}},
\]
where $r$ ranges over the candidate ratios above. This expresses the penalty in Appendix~\ref{ex:matrix-completion} relative to the smallest value for which the low-rank component is zero. The corresponding singular-value threshold is $r\|\mathbf R\|_{\mathrm{op}}$. We use the same rule for observed rates and latent log rates.

For GSC, we jointly estimate the policy coefficients, unit and period effects, and interactive factors using never-treated controls. We then estimate each treated unit's intercept and loadings from its pretreatment outcomes, holding the common coefficients and factors fixed. We compare specifications by averaging squared prediction errors within each fold and then giving the folds equal weight.

To compare point predictions, we shift the six adoption schedules three to six years earlier and predict the resulting withheld outcomes. For predictive calibration, we also assign the staggered schedule to untreated states in twenty additional holdouts. The two sets receive equal total weight, with equal weights within each set.

For GSC fitted to observed rates, we use the parametric residual bootstrap of \citet{xu2017generalized}, implemented in \texttt{fect}, with 2,000 replicates and the selected rank fixed. Resampling preserves residual trajectories and uses out-of-sample control prediction errors for treated units with the corresponding pretreatment lengths. We refit the same GSC specification in each replicate. Normal confidence limits use bootstrap standard errors for the average effect over twenty treated state-years or at the relevant event time.

For comparison in Figures~\ref{fig:pcpa-cannabis-event-study} and~\ref{fig:pcpa-cannabis-counterfactual}, we also estimate effects by imputation from additive models fitted to the untreated observations. We use a linear model for observed rates in (a) and the same Poisson model with population exposure in (b)--(c). Both adjust for the four policy covariates and unit and period effects, without unit-by-period interactions. The Poisson predictions are response-scale posterior means from one fit. We aggregate the predictions and observed-minus-predicted contrasts using the same event-time definitions and weights as for GSC.

\subsubsection{Results}

Cross-validation selects GSC for all three inputs, with rank two for observed rates and rank three for both Poisson specifications. Cross-validated RMSE is $1.91$ for observed rates, $1.88$ for the nonspatial Poisson model, and $1.66$ for the spatial model (Table~\ref{tab:pcpa-empirical-summary}). The posterior median spatial share is $0.38$, with $95\%$ credible interval $[0.25,0.54]$ (Table~\ref{tab:pcpa-empirical-dependence}).

\begin{figure}[!tbp]
\centering
\caption{Observed and counterfactual opioid-mortality rates in the recreational-cannabis evaluation.}
\includegraphics[width=\textwidth]{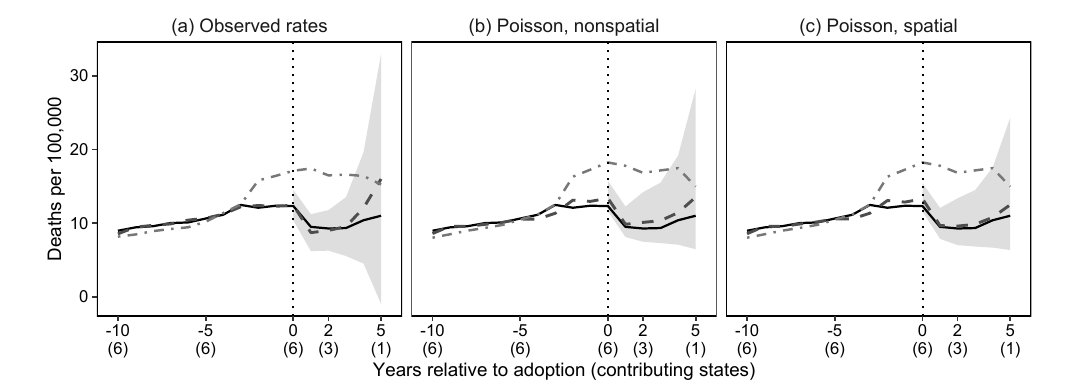}
\figurenote{Observed opioid-mortality rates (solid), imputed untreated rates from GSC (dark dashed), and imputed untreated rates from additive models (gray dot-dashed), plotted by years relative to legalization. GSC uses observed rates in (a), posterior draws from a nonspatial Poisson model in (b), and posterior draws from a spatial Poisson model in (c). The additive models include four policy covariates and unit and period effects but no unit-by-period interactions; they use a linear model on observed rates in (a) and the same additive Poisson fit in (b)--(c). Rates are deaths per 100,000 and are averaged equally across contributing states; numbers of contributing states appear below the axis. Shaded bands apply only to GSC and show pointwise 95\% residual-bootstrap confidence intervals in (a) and posterior predictive intervals in (b)--(c).}
\label{fig:pcpa-cannabis-trajectories}
\label{fig:pcpa-cannabis-counterfactual}
\end{figure}

The estimated average effect is $-0.25$ deaths per $100{,}000$ using observed rates, with bootstrap confidence interval $[-2.79,2.30]$. The nonspatial and spatial Poisson analyses give predictive medians of $-0.95$ and $-0.62$, with predictive intervals of $[-4.20,1.04]$ and $[-3.53,1.36]$, respectively. The spatial model gives somewhat narrower predictive intervals than the nonspatial model. The event-time estimates and counterfactual trajectories are shown in Figures~\ref{fig:pcpa-cannabis-event-study} and~\ref{fig:pcpa-cannabis-counterfactual}.

\FloatBarrier

\subsection{Vision Zero and traffic fatalities}
\label{subsec:pcpa-vision-zero}
\label{subsec:vision-zero}

We revisit Sweden's Vision Zero road-safety policy \citep{bonander2021flexible}. Ordinary synthetic controls fit Sweden's low pretreatment fatality rates poorly, motivating ridge augmentation. We compare augmented synthetic controls on observed rates with the same estimator applied to latent log rates from a Poisson model.

\subsubsection{Data and analysis}

The Swedish Parliament adopted Vision Zero in 1997 as the guiding framework for national road-safety policy \citep{bonander2021flexible}. We use annual traffic fatalities for Sweden and nine controls from 1970 through 2015, with passenger travel as exposure. We fit SCM and ASCM to observed rates and combine ASCM with the nonspatial Poisson model. Effects are expressed as fatalities per billion passenger-kilometers and averaged over 1997--2015.

For ASCM \citep{ben2021augmented}, we evaluate 17 penalty ratios,
\[
r\in\left\{10^a:a=-4,-3.5,\ldots,4\right\}.
\]
We use the standardization and penalty scaling in Appendix~\ref{app:pcpa-ascm-implementation}, recalculating the penalty for each input panel. We select the ratio by cross-validation, whereas it was fixed in the Poisson experiment.

To compare point predictions, we use fourteen three-year holdouts beginning annually from 1981 through 1994 and withhold all subsequent Swedish outcomes. Predictive calibration instead matches the nineteen-year post-intervention average. We use nine control-country holdouts beginning in 1997 and twenty-seven beginning in 1978, 1984, or 1990. Each evaluates predictions over nineteen years, with later outcomes for that country also removed. Holdouts beginning in 1997 and those beginning earlier receive equal total weight, corresponding to weights of $1/18$ and $1/54$ per holdout, respectively.

For SCM and ASCM applied to observed rates, we construct pointwise conformal intervals using time-shift permutations following \citet{chernozhukov2021conformal}. We use their tuning choice $q=1$ and a 201-point evaluation grid, enumerating the cyclic time shifts.

\subsubsection{Results}

Augmentation improves prediction relative to ordinary SCM (Figure~\ref{fig:pcpa-vision-zero}). Cross-validated RMSE is $2.88$ for SCM and $0.91$ for ASCM on observed rates, and $0.52$ for ASCM with the Poisson model. The selected penalty ratios are $10$ and $0.1$ for the two augmented analyses, respectively.

The estimated average effect is $-1.39$ fatalities per billion passenger-kilometers under SCM and $0.15$ under ASCM on observed rates. The Poisson analysis gives a predictive median of $0.25$, with $95\%$ predictive interval $[-3.56,1.96]$ (Table~\ref{tab:pcpa-empirical-summary}). Most of the change in the estimated effect comes with augmentation; the Poisson model improves prediction further without a comparable change in the point estimate.

\subsection{Sensitivity of predictive inference}
\label{app:pcpa-empirical-sensitivity}
\label{app:empirical-predictive-sensitivity}

We assess sensitivity to the innovation scales by adding independent and joint Gaussian candidates at four times the reference standard deviation. We retain the fitted probabilistic models, causal estimators, and validation designs, and re-estimate only the predictive mixture weights.

The cigarette-sales and opioid-mortality results change little (Table~\ref{tab:pcpa-empirical-sensitivity}, Panel A). For Vision Zero, the predictive median changes from $0.25$ to $0.29$, while the $95\%$ interval widens from $[-3.56,1.96]$ to $[-4.92,2.09]$, mainly toward larger reductions in fatalities.

We also assess the influence of the Croatian validation holdouts by omitting them when estimating the five-candidate mixture weights. Croatia remains in Sweden's control pool, and the candidate predictions for Sweden are unchanged. We retain equal total weight for holdouts beginning in 1997 and those beginning earlier. The predictive median is $0.26$ and the $95\%$ interval is $[-3.56,1.96]$, close to the result using all holdouts (Panel B).

\begin{table}[!tbp]
\centering
\begin{threeparttable}
\renewcommand{\TPTminimum}{\textwidth}
\caption{Sensitivity of treatment-effect estimates and predictive intervals.}
\label{tab:pcpa-sensitivity}
\label{tab:pcpa-empirical-sensitivity}
\label{tab:empirical-predictive-sensitivity}
\small\setstretch{1}
\setlength{\tabcolsep}{0pt}
\begin{tabular*}{\linewidth}{@{}>{\raggedright\arraybackslash}p{\dimexpr\linewidth-358pt\relax}
@{\hspace{10pt}}>{\raggedright\arraybackslash}p{72pt}
@{\hspace{10pt}}>{\centering\arraybackslash}p{128pt}
@{\hspace{10pt}}>{\centering\arraybackslash}p{128pt}@{}}
\toprule
\multicolumn{4}{l}{\textit{Panel A. Sensitivity to the range of innovation scales}} \\
\addlinespace[3pt]
Application & Probabilistic model & Five candidates & Seven candidates \\
\midrule
California & Nonspatial & $-18.8\;[-42.5,\,5.3]$ & $-18.7\;[-42.8,\,5.5]$ \\
California & Spatial & $-18.7\;[-40.7,\,3.9]$ & $-18.7\;[-40.6,\,3.9]$ \\
Cannabis & Nonspatial & $-0.95\;[-4.20,\,1.04]$ & $-0.95\;[-4.20,\,1.04]$ \\
Cannabis & Spatial & $-0.62\;[-3.53,\,1.36]$ & $-0.62\;[-3.53,\,1.36]$ \\
Vision Zero & Nonspatial & $0.25\;[-3.56,\,1.96]$ & $0.29\;[-4.92,\,2.09]$ \\
\bottomrule
\end{tabular*}
\par\medskip
\begin{tabular*}{\linewidth}{@{}>{\raggedright\arraybackslash}p{\dimexpr\linewidth-160pt\relax}
@{\hspace{10pt}}S[table-format=1.2,table-column-width=46pt]
@{\hspace{10pt}}>{\centering\arraybackslash}p{94pt}@{}}
\multicolumn{3}{l}{\textit{Panel B. Vision Zero sensitivity to the validation sample}} \\
\addlinespace[3pt]
Validation holdouts & {Median} & 95\% interval \\
\midrule
All holdouts & 0.25 & $[-3.56,\,1.96]$ \\
Excluding Croatia as validation target & 0.26 & $[-3.56,\,1.96]$ \\
\bottomrule
\end{tabular*}
\begin{tablenotes}[flushleft]
\footnotesize
\item Notes: Predictive medians with 95\% predictive intervals in the application units of Table~\ref{tab:pcpa-empirical-summary}. Nonspatial and Spatial identify the corresponding probabilistic specifications. Panel A expands the five-candidate mixture with independent and joint Gaussian innovations at four times the reference standard deviation, holding the fitted models and validation design fixed. Panel B omits Croatia as a validation target when estimating the five-candidate weights but retains it in Sweden's control pool. Only predictive mixture weights are re-estimated in these comparisons.
\end{tablenotes}
\end{threeparttable}
\end{table}

\FloatBarrier

\end{document}